\documentclass[sigconf]{acmart}

\usepackage{multirow}
\usepackage{color}
\usepackage{colortbl}
\usepackage{stfloats}
\usepackage{subcaption}
\usepackage{enumitem}
\usepackage{graphicx}
\usepackage{subcaption}
\usepackage{tabularx}
\usepackage{ifthen}

\usepackage{graphicx}
\usepackage{url}
\usepackage{makecell}
\graphicspath{ {./images/} }

\usepackage{hwemoji} 
\usepackage[utf8]{inputenc}

\usepackage{longfbox}

\makeatletter
\newdimen\@tempdimd
\makeatother

\newcommand{\sourcecode}{\urlstyle{tt}\url{https://naver-ai.github.io/elmi}}

\definecolor{quotebackground}{HTML}{EFEFEF}
\definecolor{tableheader}{HTML}{EFEFEF}
\definecolor{tableheaderlight}{HTML}{f8f8f8}
\definecolor{tablegrayline}{HTML}{e0e0e0}

\newcommand{\eg}{\textit{e.g.}}
\newcommand{\ie}{\textit{i.e.}}
\newcommand{\cf}{\textit{c.f.}}
\newcommand{\etal}{\textit{et al.}}

\newcommand{\revised}[1]{#1} 

\newcommand{\delete}[1]{} 

\newcommand{\cameraready}[1]{#1} 

\newcommand{\circledigit}[1]{\textbf{\normalsize{\textsf{\textcircled{\footnotesize{#1}}}}}}

\newcommand{\ipstart}[1]{\vspace{1mm} \noindent{\textbf{\textit{#1.}}}}

\newcommand{\bpstart}[1]{\vspace{1mm} \noindent{\textbf{#1}}}

\newcommand{\meaningicon}{\raisebox{-1pt}{\includegraphics[width=9pt]{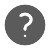}} }
\newcommand{\glossingicon}{\raisebox{-1pt}{\includegraphics[width=9pt]{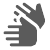}} }
\newcommand{\emotingicon}{\raisebox{-1pt}{\includegraphics[width=9pt]{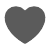}} }
\newcommand{\timingicon}{\raisebox{-1pt}{\includegraphics[width=9pt]{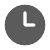}} }

\newcommand{\assignedsong}{Song\raisebox{1pt}{{\tiny{$^{Assigned}$}}}}
\newcommand{\selectedsong}{Song\raisebox{1pt}{{\tiny{$^{Selected}$}}}}

\usepackage{newfloat}
\DeclareFloatingEnvironment[fileext=fml,placement={H},name=Dialogue]{dialogue}
\newenvironment{quotetable}{
\hfill\break
\noindent
    \small\sffamily\centering
    \def\arraystretch{1.15}\setlength{\tabcolsep}{0.25em}
    \tabularx{\columnwidth}{cm{0.85\columnwidth}}
    \hline}
{
    \arrayrulecolor{black}\hline
    \endtabularx
}

\newcommand{\quotebotline}[1]{\textbf{ELMI} & \textit{#1}\\\arrayrulecolor{tablegrayline}\hline}

\newcommand{\quoteuserline}[2]{\textbf{#1} & \textit{#2}\\\arrayrulecolor{tablegrayline}\hline}

\newfboxstyle{boxparam}{padding=1.5pt, padding-left=2pt, padding-right=2pt, height=7.5pt, background-color=lightgray, border-radius=2pt, border-right-width=1pt, border-bottom-width=1pt, border-color=gray}

\newcommand{\rectwrapsmall}[2]{\lfbox[boxparam, border-radius=0pt, padding-left=2pt, padding-right=2pt, height=5.5pt, border-width=0pt, background-color=#1]{\sffamily{\textcolor{white}{#2}}}}

\newcommand{\blackrectsmall}[1]{\rectwrapsmall{darkgray}{#1}}

\definecolor{deafcolor}{HTML}{fad2dd}
\definecolor{hhcolor}{HTML}{fad2dd}
\definecolor{hearingcolor}{HTML}{cde0ec}

\newcommand{\deafwrap}[1]{\lfbox[boxparam, border-width=0pt, background-color=deafcolor]{#1}}

\newcommand{\hhwrap}[1]{\lfbox[boxparam, border-width=0pt, background-color=hhcolor]{#1}}
\newcommand{\hearingwrap}[1]{\lfbox[boxparam, border-width=0pt, background-color=hearingcolor]{#1}}

\newcommand{\p}[1]{%
    \ifthenelse{\numexpr#1<4}{%
        \deafwrap{P#1\raisebox{1pt}{{\tiny{$^{d/D}$}}}}}{%
        \ifthenelse{\numexpr#1<7}{%
            \hhwrap{P#1\raisebox{1pt}{{\tiny{$^{d/D}$}}}}}{%
            \hearingwrap{P#1\raisebox{1pt}{{\tiny{$^{Hg}$}}}}}%
    }%
}

\newcommand{\s}[1]{%
    \ifthenelse{\numexpr#1<3}{%
        \deafwrap{S#1\raisebox{1pt}{{\tiny{$^{d/D}$}}}}}{%
        \ifthenelse{\numexpr#1<6}{%
            \hhwrap{S#1\raisebox{1pt}{{\tiny{$^{d/D}$}}}}}{%
            \hearingwrap{S#1\raisebox{1pt}{{\tiny{$^{Hg}$}}}}}%
    }%
}

\newcommand{\emojismile}{\raisebox{-2pt}{\includegraphics[width=9pt]{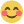}} }
\newcommand{\emojidanger}{\raisebox{-2pt}{\includegraphics[width=9pt]{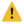}} }

\copyrightyear{2025}
\acmYear{2025}
\setcopyright{rightsretained}
\acmConference[CHI ’25]{CHI Conference on Human Factors in Computing Systems}{April 26-May 1, 2025}{Yokohama, Japan}
\acmBooktitle{CHI Conference on Human Factors in Computing Systems (CHI ’25), April 26-May 1, 2025, Yokohama, Japan}\acmDOI{10.1145/3706598.3713973}
\acmISBN{979-8-4007-1394-1/25/04}

\begin{document}

\title{ELMI: Interactive and Intelligent Sign Language Translation of Lyrics for Song Signing}

\author{Suhyeon Yoo}
\authornote{Suhyeon Yoo conducted this work as a research intern at NAVER AI Lab.}
\orcid{0000-0001-6098-9947}
\affiliation{%
  \institution{University of Toronto}
  \country{Toronto, Canada}
}
\email{suhyeon.yoo@mail.utoronto.ca}

\author{Khai N. Truong}
\orcid{0000-0003-0774-5964}
\affiliation{%
  \institution{University of Toronto}
  \country{Toronto, Canada}
}
\email{khai@cs.toronto.edu}

\author{Young-Ho Kim}
\orcid{0000-0002-2681-2774}
\affiliation{%
  \institution{NAVER AI Lab}
  \country{Republic of Korea}
}
\email{yghokim@younghokim.net}

\begin{abstract}
d/Deaf and hearing song-signers have become prevalent across video-sharing platforms, but translating songs into sign language remains cumbersome and inaccessible.
Our formative study revealed the challenges song-signers face, including semantic, syntactic, expressive, and rhythmic considerations in translations.
We present ELMI, an accessible song-signing tool that assists in translating lyrics into sign language. ELMI enables users to edit glosses line-by-line, with real-time synced lyric and music video snippets. Users can also chat with a large language model-driven AI to discuss meaning, glossing, emoting, and timing.
Through an exploratory study with 13 song-signers, we examined how ELMI facilitates their workflows and how song-signers leverage and receive an LLM-driven chat for translation.
Participants successfully adopted ELMI to song-signing, with active discussions throughout. They also reported improved confidence and independence in their translations, finding ELMI  encouraging, constructive, and informative.
We discuss research and design implications for accessible and culturally sensitive song-signing translation tools.

\end{abstract}



\begin{teaserfigure}
  \includegraphics[width=\textwidth]{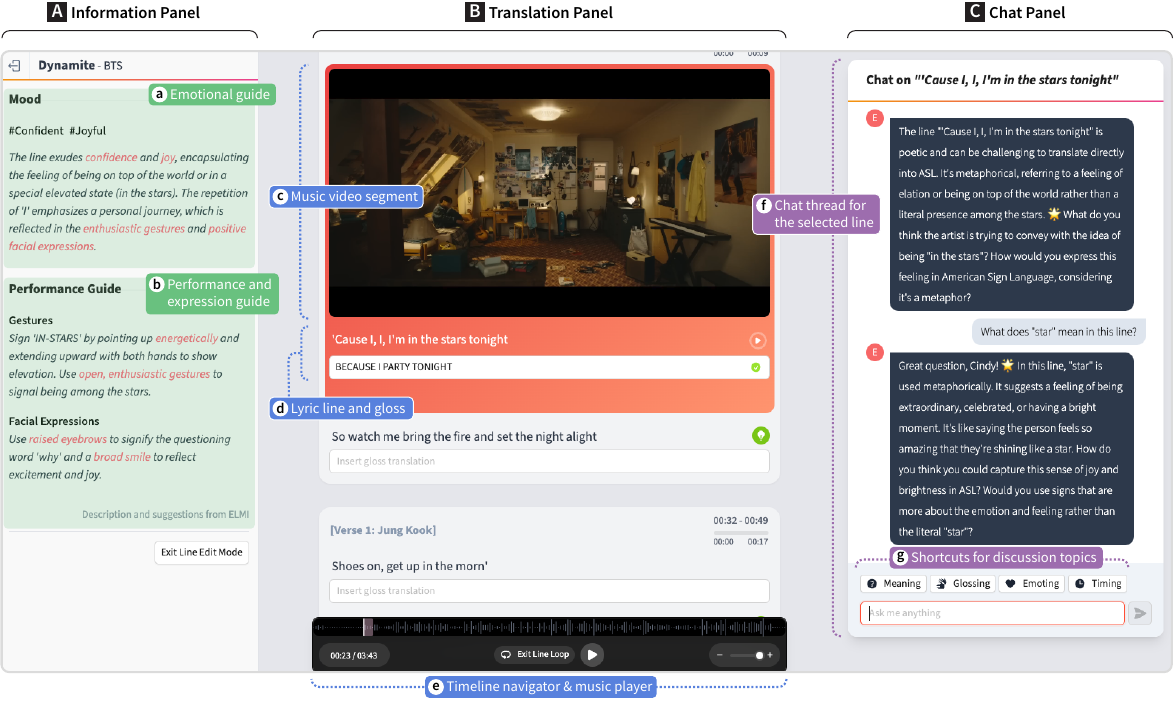}
  \caption{ELMI's translation interface, supporting line-by-line translation of lyrics into sign language. 
  The Information Panel~\blackrectsmall{A} provides background information about the song or AI-generated description on mood and performance guide for the selected lyric line. \revised{The emotional guide~\circledigit{a} suggests hashtags that capture the mood of each line, while the performance and expression guide~\circledigit{b} offers recommendations for bodily and facial expressions to effectively convey these emotions.}
  On the Translation Panel~\blackrectsmall{B}, the user can write gloss (text representation of sign language) for the selected lyric line~\circledigit{\revised{d}}, looping the music video segment~\circledigit{\revised{c}}. The Chat Panel~\blackrectsmall{C} provides a discussion channel with a large language model-driven chatbot regarding the selected line~\circledigit{\revised{f}}. For the paper figures, we used the official music video of BTS's Dynamite, publicly available at Youtube~(\url{https://www.youtube.com/watch?v=gdZLi9oWNZg}).}
  \Description{This figure illustrates the main screen of ELMI's translation interface, which is divided into three panels. The Information Panel (A), positioned on the left side of the screen, provides background details about the song or AI-generated descriptions of the mood (a) with hashtags and performance guidance (b) with gestures and facial expressions for the selected lyric line. The Translation Panel (B), located in the center, allows the user to write glosses (text representations of sign language) and loop segments of the music video. The video segment (c) is displayed at the top, followed by the lyric line and gloss (d) in the middle, with the timeline navigator and music player (e) at the bottom. On the right side of the screen is the Chat Panel (C), which facilitates a discussion with an AI-driven chatbot about the selected lyric line. The chat thread (f) includes shortcut buttons for four discussion topics: meaning, glossing, emotion, and timing (g).}
  \label{fig:teaser}
\end{teaserfigure}

\maketitle

\section{Introduction}
Song-signing\footnote{Also known as sign singing, signed song, sign singing, or karaoke signing \cite{cripps2019signed}. \revised{Song-signing involves translations/interpretations of pre-existing songs into a signed language, original pieces composed in sign language, or arrangements of signs to specific beats known as percussion signing (\eg{}, ``Bison Song'' by Dorothy Miles)~\cite{maler2013songs, bahan2006face}. The signed songs can be presented as recorded videos (\eg{}, YouTube performances) or as live presentations (\eg{}, concerts or stage performances)~\cite{maler2013songs, bahan2006face}.}} is a performing art, encompassing not only translating spoken lyrics to sign language but also conveying musical elements through facial expressions and bodily movements~\cite{bahan2006face, maler2013songs}. 
d/Deaf\footnote{In research, the acronym DHH (Deaf and Hard of Hearing) is often used to refer to individuals with varying degrees of hearing loss \cite{guardino2016deafness}. However, in this study, we use ``d/Deaf'' to emphasize cultural identity. ``deaf'' (with a lowercase `d') refers to the physical condition of hearing loss. These individuals may use hearing aids, cochlear implants, or rely on lip reading. On the other hand, ``Deaf'' (with an uppercase `D') refers to individuals who identify as part of the Deaf community, share a common language (such as ASL), and embrace Deaf culture and identity \cite{skelton2003feels}.} and hearing song-signers perform in visually dynamic forms, blending the rhythm and emotion of songs with the physical expressiveness of sign language \cite{fisher2021embodied}. 
Translation quality is crucial, as it enhances the message of the song by incorporating emotional tone, timing, and cultural nuances, going beyond the literal meaning of the words~\cite{maler2013songs, bahan2006face, cripps2019signed}.
This requires a deep understanding of both spoken and sign languages, ensuring that the final signed performance is both accurate and resonant~\cite{yoo2023understanding, snoddon2024signing}.

However, translating song lyrics into signs presents challenges. Often, sign language may not have the equivalent vocabulary for some terms from the spoken language, necessitating the use of fingerspelling or creative alternatives \cite{yoo2023understanding}. Furthermore, since the syntax and modality of two languages are distinct, the \textit{glosses}\footnote{Gloss refers to a written representation of a signed language, using words from a spoken language (like English) to indicate the meaning of individual signs. Glossing helps translate the structure and meaning of a signed language but does not fully capture nuances such as facial expressions, which are crucial elements of communication \cite{supalla2017american}.} should be adjusted to fit the rhythm of the music and the signing space performing sign language~\cite{snoddon2024signing, yoo2023understanding}.
While past research has explored ways to make music more accessible for d/Deaf individuals, much of this work has focused on music perception \cite{karam2010emoti, mchugh2021towards} and creation \cite{petry2018supporting, ohshiro2022people} rather than performance.
Research on the artistic interpretation of sign language remains sparse, and prior work predominantly focused on communication accessibility \cite{rui2022online}. Only a few projects have recently begun to understand song-signing~\cite{yoo2023understanding}. For example, Yoo \etal{} identified tensions between d/Deaf and hearing song-signers and the different barriers they encounter in the process of song-signing. Still, little effort has been put into designing accessible systems for song-signing.

In this work, we aim to design a system that supports \revised{translation of song lyrics and gloss creation for} song-signing in a more accessible manner. 
To better understand the current song-signing translation process and challenges, we conducted formative interviews with eight song-signers 
(five d/Deaf and three hearing). The study revealed 
four major challenges of song-signing in the translation process: 
semantic translation (\eg{}, understanding the meaning of the lyrics), syntactic translation (\eg{}, selecting appropriate signs) \footnote{\cameraready{In this work, we use the term \textit{syntactic translation} to indicate the process of translating lyrics into glosses while ensuring that the sign order aligns with ASL grammar and maintains the musical timing~\cite{yoo2023understanding}.}}, expressive translation (\eg{}, conveying emotion), and rhythmic translation (\eg{}, aligning the timing and length of signs with the music). 
As a result, many song-signers rely on external resources such as social media videos and discussions with other interpreters to improve their work.


To address these challenges, we designed and developed \textbf{ELMI} (Explore Lyrics and Music Interactively; \autoref{fig:teaser}), a web-based accessible song-signing tool that facilitates lyric-sign translation. ELMI features a line-by-line focus, allowing the user to input gloss for each line and play corresponding video segments. For a d/Deaf user, ELMI offers rich visual feedback to convey music, such as a music video player with a karaoke-style timing indicator. Inspired by the generative capabilities of large language models (LLMs), ELMI leverages LLMs to provide various AI-driven guidance. For example, to aid in familiarization with the song, ELMI provides AI-generated annotations of the lyric lines, such as emotional tone and performance guidance. For each lyric line, the user can chat with ELMI to discuss key topics of gloss creation, including meaning, glossing, emotion, and timing.

We evaluated ELMI through an exploratory study with 13 song-signers 
(\revised{6 d/Deaf}, and 7 hearing), who used the tool to create glosses for two songs.
ELMI \delete{significantly improved} \revised{facilitated the} translation \delete{quality} \revised{process} by producing \delete{more} nuanced and contextually accurate glosses.
Participants reported increased confidence and independence in their translation process through their discussions with ELMI.
Both d/Deaf and hearing participants found glossing and meaning-related discussions useful while hearing favored emoting, and d/Deaf valued timing assistance.
ELMI streamlined decision-making, providing all required resources in one place. Participants perceived ELMI as encouraging, critical, and informative in giving them full control and ownership over their glosses.
All participants expressed interest in incorporating ELMI into their workflow, as it helped them structure their process. 
The key contributions of this work are threefold:
\begin{enumerate}[leftmargin=*, itemsep=4pt, topsep=2pt]
\item The design and implementation of ELMI, an AI-infused song-signing tool that supports translating lyrics into sign language. The source code of ELMI is publicly available at \sourcecode{}.
\item The empirical understanding of how song-signers perceive and interact with ELMI, creatively producing glossing through discussion with the tool.
\item Implications for future research and systems on acceptability and accountability for artistic sign language translation, highlighting the need to respect cultural contexts and maintain artistic integrity.
\end{enumerate}
\section{Related Work}
In this section, we cover related work in three parts: (1) song-signing, (2) technologies for automated sign language translation, and (3) lyric translation technologies. 

\subsection{Song-signing}
Song-signing is an art form rooted in Deaf culture~\cite{maler2013songs}. Song-signers convey not only the lyrics but also capture musical elements such as rhythm, emotion, and instruments through facial expressions and bodily gestures \cite{yoo2023understanding}. Song-singing performance is crucial in making music accessible to d/Deaf individuals by providing an augmented visual representation \cite{yoo2023understanding}. It has also proved to be more meaningful and authentic to those fluent in sign language, offering a genuine and enjoyable experience \cite{crippsunderstanding}. 

The core activity of song-signing---trans-languaging from spoken language to sign language---requires a high degree of creativity, encompassing the translation of lyrics or the composition of original songs in sign language, where performers manipulate signs and signing space to convey musical features \cite{maler2013songs}. 
This practice involves complex interactions between music, lyrics, and sign language, requiring artists to navigate across languages and modalities \cite{snoddon2024signing}. 
Yoo \etal{} proposed three layers in song-signing translation: technical translation, understanding the lyrics and matching gloss with the music; artistic translation, editing the gloss to convey non-lyrical elements like pitch, rhythm, and volume; and cultural translation, appreciating of Deaf culture and language \cite{yoo2023understanding}


Recent studies on song-signing have identified several challenges, including the difficulty of translating lyrics into sign language, conveying musical features, and addressing diverse linguistic repertoires \cite{yoo2023understanding, maler2013songs}. 
Song-signing involves navigating between visual and auditory channels and integrating music, lyrics, and sign language \cite{snoddon2024signing}. 
One significant challenge is that many song lyrics carry symbolic meanings rather than straightforward messages. When translated into sign language, the interpretation is shaped by the translator, which can lead to a disconnect between the original intent of the writer and the signed interpretation \cite{crippsunderstanding}. 
Additionally, song-signing can appear weak if the translator struggles with timing, particularly when trying to produce musical American Sign Language (ASL) within the time constraints of English songs \cite{crippsunderstanding}. 
Although these challenges underscore the need for culturally responsive approaches to music accessibility \cite{robinson2022knowing, yoo2023understanding}, research on how technology can support song-signing translation remains limited.


\subsection{Automated Sign Language Translation}

Sign language is the primary mode of communication for d/Deaf individuals \cite{saunders2023different}, and research in this area is crucial for facilitating communication between Deaf communities and others \cite{papastratis2021artificial}. As one dominant line of efforts that bridge the gap between spoken and sign languages, ample research in machine learning and accessibility has presented automated models that convert signed sentences into spoken or written language and vice versa~\cite{papastratis2021artificial, liang2023sign, nunez2023survey, desai2024systemic}. As sign language is carried over bodily movements, video recognition or generation is required for ideal end-to-end translation~\cite{liang2023sign}. To make the task straightforward, gloss text has often been used as a representation medium of sign language, and a majority of machine learning approaches tackled gloss-spoken language translation~\cite{nunez2023survey, desai2024systemic}. For example, researchers proposed gloss-based neural machine translation models~(\eg{},~\cite{moryossef2021data,amin2021sign,camgoz2020sign,cao2022explore}), Recognizing the loss of partial linguistic cues when using gloss as a representation of sign language~\cite{desai2024systemic, muller2023considerations}, recent work leveraged large language models (LLMs)~\cite{openai2023gpt4} to implement gloss-free sign language translation in combination with vision models~\cite{gong2024llms, wong2024sign2gpt}. 

Despite the growing body of automated sign language translation technologies, existing models focus on communication and mapping the meanings, trained on datasets in general contexts such as word-level sign lexicons~(\eg{}, \cite{li2020word, Vassilis2008ASLLVD}), news and weather forecast~(\eg{}, \cite{camgoz2018neural}), and daily conversations~(\eg{},~\cite{vaezijoze2019msasl}). 
In other words, they are not designed with lyric translation in mind, which critically involves considerations of timing and poetic interpretation~\cite{yoo2023understanding}. Hence, lyric-to-sign-language translation should incorporate users in the loop, rather than being carried over a fully-automated approach. In this work, we support lyric translation driven by song-signers, while facilitating the translation process with AI-driven guidance and discussions.

\subsection{Lyric Translation and Tools}


Emotion and timing are crucial in translating lyrics due to their significant impact on the overall musical experience. Translators must navigate the complex interplay between language, music, and cultural nuances to maintain the emotional resonance of the original song \cite{tambunan2024implicit}. This translation process requires careful consideration of both implicit and explicit information, as well as the emotional content of the source text \cite{hubscher2017translation}. Franzon describes the strategic choices made by translators and lyricists in achieving `singability,' the musico-verbal fit of text to music—by balancing prosodic, poetic, and semantic layers \cite{franzon2008choices}. The emotional intelligence of translators also plays a pivotal role in their decision-making and problem-solving during translation: understanding of linguistic, cultural, and emotional subtleties to effectively bridge linguistic and cultural boundaries in music \cite{hubscher2017translation, tambunan2024implicit}.

New tools for lyric translation and interpretation have been introduced. Statistical Machine Translation was applied to an automated lyric annotation system to clarify complex terminology and abstract concepts using the Genius dataset \cite{sterckx2017break}. 
Semi-automatic lyric generation using context-free grammar has shown promise, with generated lyrics sometimes indistinguishable from those written by humans \cite{pudaruth2014automated}. 
Recent research has explored the potential of large language models (LLMs) in supporting lyric interpretation and song understanding. For instance, BART-fusion combines a pre-trained language model with an audio encoder to generate lyric interpretations, showing improved performance with the integration of audio information \cite{zhang2022interpreting}. 
These studies demonstrate the versatility of LLMs in understanding and generating music-related content, highlighting their potential to enhance interpretability across various domains.


Despite these technological advancements, the translation of lyrics continues to pose significant challenges. Translators must capture stylistic elements, cultural references, and semantic nuances while maintaining fluency and creativity \cite{andayani2023inquiry}. The ambiguous nature of lyrics, along with the use of slang, jargon, and the need to preserve rhyme and repetition, adds to the complexity of this task \cite{apter2016translating}. Moreover, translators face difficulties related to rhythm, syllable count, vocal burden, and the need to adapt cultural nuances across languages \cite{apter2016translating}. Cultural differences often necessitate adaptations and retranslations to accurately convey meaning. Translating idioms within lyrics, in particular, requires a deep understanding of both the source and target languages to ensure the intended meanings are preserved \cite{andayani2023inquiry}. Addressing these challenges requires interdisciplinary approaches that combine linguistics, artificial intelligence, and cultural studies. 
This work extends the body of research on lyric translation within the context of song-signing, focusing on artistic and cultural dimensions in the translation process.

\section{Formative Study}
To understand the current translation process and challenges in song-signing, as well as to explore the applicability of LLM on song-signing translation,  we conducted semi-structured interviews with song-signers. The study protocol was approved by the university research ethics board. 

\subsection{Methods}

\ipstart{Participants} 
We recruited eight song-signers (\revised{\s{1}--\s{8}; five d/Deaf and three hearing in order. See Appendix~\ref{appendix:formative:demographic} for detailed demographic information.}) comprising five males and three females, with ages ranging from 25 to 67 years old. Our inclusion criteria were song-signers who are 18 years or older, proficient in speaking and writing English, and have experience in song-signing within the past year. Participants were recruited through emails, social media platforms (e.g., Facebook, YouTube), and the snowball sampling method.
The majority of participants ($N=5$) used American Sign Language (ASL) at expert and native levels, while others also utilized Pidgin Signed English (PSE). 
On average, participants have been involved in song-signing for 9.3 years, spanning from 4 to 17 years. 
While some participants used tools such as dictionaries for translation, the use of LLM-driven tools like ChatGPT was minimal, with most participants relying on personal knowledge and community resources.

\begin{figure}[t]
    \centering
    \begin{subfigure}[b]{\linewidth}
        \centering
        \includegraphics[width=\textwidth]{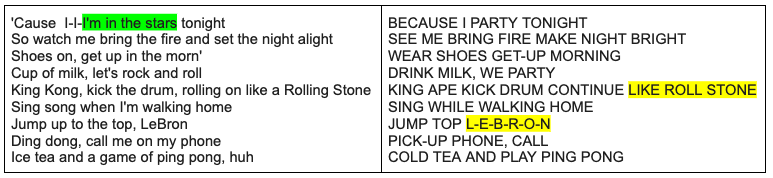}
        \caption{Excerpt of glosses from \s{6}}
        \vspace{5mm}
    \end{subfigure}
    \begin{subfigure}[b]{\linewidth}
        \centering
        \includegraphics[width=0.8\textwidth]{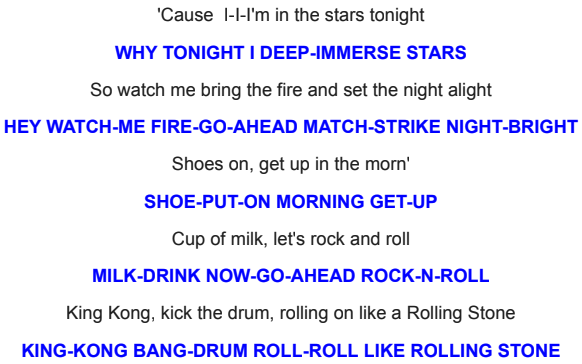}
        \caption{Excerpt of glosses from \s{7}}
    \end{subfigure}
    \caption{Example Glossing for ``BTS - Dynamite.'' Song-signers created glosses line-by-line, writing ASL glosses corresponding to ENG lyrics. }
    \Description{"This figure presents example glosses from song-signers S6 and S7 for the BTS song "Dynamite," created line-by-line. The top subfigure (a) shows S6's gloss, presented in a 1x2 table format, with the English lyrics on the left and the ASL gloss on the right. The below subfigure (b) shows S7's gloss, where the English lyrics are displayed in black, and the corresponding ASL gloss is written underneath in blue uppercase letters."}
    \label{gloss}
\end{figure}


\ipstart{Procedure} 
We invited each participant to Zoom sessions which lasted between 50 to 70 minutes. We utilized Zoom chats, closed captioning, and sign language interpreters for d/Deaf participants who preferred those options\delete{rather than spoken language}. \revised{Participants’ signed responses were interpreted in real-time by certified ASL interpreters, whose spoken translations were then transcribed into English.}
%
The interviews covered (1) the current work process and practice method for translating English lyrics to sign language, (2) factors and considerations influencing translation, and (3) challenges and workarounds in translation. 
To collect participants' preliminary feedback on using LLMs in song-signing, we asked them to try ChatGPT~\cite{OpenAI2024ChatGPT} while translating a snippet of the example song "Dynamite" by BTS (See~\autoref{gloss}). 
\revised{During this task, participants were encouraged to ask ChatGPT any open-ended questions that came to their mind regarding various aspects of the translation process. Participants engaged ChatGPT in diverse ways,}
such as understanding the context of the song, interpreting specific lyrics, seeking confirmation on glossing and interpretation, finding videos of performances or song-signings, and providing images or videos of example signs. 

\ipstart{Analysis} 
We recorded the video call sessions and transcribed them for analysis. 
\revised{Using a thematic analysis framework \cite{braun2012thematic, braun2019reflecting}, two researchers independently coded the transcripts to identify key ideas and grouped similar codes into categories. Through multiple rounds of discussion, the researchers refined and reviewed themes to ensure they accurately reflected the data and aligned with the research objectives.}

\subsection{Lyric Translation Practice}
\subsubsection{Familiarization}
Participants begin by listening to the music and watching music videos to understand the context of the song's lyrics.
Rather than focusing on a word-for-word translation, song-signers aim for \textit{conceptual accuracy}, ensuring that the essence and message of the song are preserved. 
During this process, an initial gloss is created that captures the general meaning of the song. 
The lyrics are then broken down into chunks and eventually ``line by line", with each segment translated into sign language, ensuring that each sign fits within the overall narrative of the song. 
Additionally, participants conducted contextual and cultural research to understand the background and significance of the song: ``\textit{I do a little bit of research why the lyric was written, why the artist chose the song, and how they sang it} (\s{1})." 

\subsubsection{Practice and Refinement} 
After creating the initial version of the gloss, participants refine it to ensure the smoothness and visual appeal of the translation. 
\s{4} remarked, ``\textit{In practice, I have an idea of what I want to sign. Sometimes I sign with the music and come across certain spots that are very tricky. Then I might go back and reevaluate if there's a simpler way to sign it.}''
Artistic elements such as rhyming were emphasized to retain the artistic integrity of the original song and make the performance visually pleasing. Two participants (\s{2}, \s{6}) noted the importance of utilizing the five parameters of ASL---handshape, location, movement, palm orientation, and non-manual markers.

Participants commonly used glosses to understand and memorize the signs. 
Participants also used glossing while performing, often setting up cameras with reference glosses with lyrics positioned nearby for visibility.
Half of the participants ($N$ = 4) used online editing tools like Google Docs to create glosses (See~\autoref{gloss}).
However, as song-signers became more experienced, they tended not to use written glosses but to practice directly with the music.

\subsection{Challenges of Song-signing Translation}
\ipstart{Semantic Translation [Meaning]}
Participants pointed out the difficulty of grasping the meaning of the lyrics. This challenge arises from the inherent complexity of the lyrics, including poetic or ambiguous expressions that can have multiple meanings ($N$ = 6). 
Accurately capturing meaning becomes even more challenging when translating concepts that are culturally or linguistically specific. \s{8} noted how accents or unfamiliar terms can lead to misinterpretations:
``\textit{[...] Another challenge is the accent. I might miss something because I don't understand a particular term.}''

\ipstart{Syntactic Translation [Glossing]} 
Since ASL relies heavily on visual elements to convey meaning, participants noted that it is particularly difficult to translate concepts that are abstract or primarily auditory in nature.
For example, three participants\delete{ (\s{1}, \s{3}, \s{6})} mentioned challenges in translating repetitive or simplistic messages in a way that feels both meaningful and visually appealing. 
Another challenge arises when English phrases do not have direct equivalents in ASL\delete{ (\s{2}, \s{3}, \s{6}, \s{8})}, requiring signers to find alternative signs or creative ways to convey the meaning. 
Also, acoustic rhymes in spoken language, which depend on sound, do not translate well into sign language.

\ipstart{Expressive Translation [Emoting]} 
Participants pointed out that beginners or those new to Deaf culture would often struggle with emotional demands in sign language due to their lack of experience and the mismatch between their own personality and the song's mood.
\s{7} remarked, ``\textit{It's a challenge, especially if you're a person who may be outwardly calm or just don't like to express yourself. Then you really have to work on that}.'' 
Certain songs, particularly those with shifting moods, demand expressive control that can be hard to master. \s{6} shared, ``\textit{I'm not the best at controlling my face. It sometimes looks awkward...Why aren't you smiling enough? One challenge is self-consciousness about your face}.''

\ipstart{Rhythmic Translation [Timing]} 
The discrepancy of modalities of spoken and signed languages imposed challenges, especially for fast-paced songs, as the rapid tempo can strain even the most skilled signers. 
Moreover, achieving synchronization with the music adds another layer of complexity, particularly for deaf signers\delete{ (\s{1}, \s{2}, \s{5})}. \s{5} described the need for extensive repetition to match the signs with the lyrics: ``\textit{It was tough, I had to repeat and repeat until I could be in sync with the music. I would repeat one sentence over 50 times to feel confident to sign as they sing.}'' This meticulous process highlights the challenge of aligning signs with the music, a task that demands significant time and effort.

\subsection{Common Strategies to Overcome Challenges}
Participants commonly looked up external resources ($N$ = 5). Many turned to online ASL dictionaries to find the appropriate signs and expressions, appreciating that the short video clips and accompanying descriptions helped clarify hand positions and sign execution. Participants also relied on search engines like Google, particularly for searching the origins and meanings of songs. 
Various types of videos---including original music videos, performance/dance videos, and song-signing videos from other song-signers---also served as a reference. All participants mentioned watching these videos helped them interpret the mood of the song and the underlying messages of the lyrics.
Participants referred to performance videos to observe facial expressions, energy levels, and choreography, which they then incorporated into their translations. Many participants also watched other song-signing videos to gain ideas and inspiration.

Notably, most participants ($N=6$) mentioned that they discussed their work with others during the translation process. They gathered feedback from a diverse group, including interpreters, Deaf \cameraready{individuals}, and young people, to refine their interpretations.
When consulting hearing individuals, participants often sought advice on visual quality and how well the lyrics matched the music. As for consulting interpreters or Deaf \cameraready{individuals}, the focus was on refining the overall translation.
Participants also consulted young people to understand contemporary slang and language trends.

\subsection{Feedback on Using LLMs for Translation}

The majority of participants shared positive views on using GPT in the translation process, recognizing its capabilities extending beyond mere literal translation, as it offers interpretations that capture the deeper meaning and emotional nuances of the original content. Additionally, GPT’s recommendations for incorporating non-manual signs were particularly impressive to participants. For example, \s{6} noted: ``\textit{My reaction was amazing as far as I can see... It asked me to use the classifier, where I found the power of that.}''
Despite its potential, participants also expressed concerns about relying on it for accurate and contextually appropriate translations because ChatGPT sometimes yielded wrong glosses or interpreted the lyrics literally, producing very long glosses.
\section{ELMI: Explore Lyrics and Music Interactively}

Informed by the formative study, we designed and developed ELMI, a novel web application that aids d/Deaf and hearing individuals in song-signing. 
\revised{Our approach leverages the potential of LLMs to facilitate the glossing process, which our formative study revealed as a critical foundation for song-signing translation. 
By facilitating AI-assisted discussions around translation nuances—such as semantic accuracy, syntactic clarity, and expressive interpretation—ELMI supports users in addressing the challenges highlighted by the song-signers in the formative study.}
In the following, we outline the design rationales for ELMI and describe the details of the interface and its implementation.

\subsection{Design Rationale}

\ipstart{DR1: Use sufficient visual \cameraready{aids} to convey emotion and timing}
Music can be appreciated not only by auditory but also by visual channels \cite{stokoe2005sign,emmorey1993visual}. 
\revised{Formative study participants also relied on lip-reading from performance videos and utilizing captions to interpret the lyrics. This aligns with prior research, where} d/Deaf individuals often lip-read and observe expressions of singers to capture musical elements \cite{breteche2018music}, \revised{while captions significantly enhance their comprehension and enjoyment \cite{mori2009seeing}.} 
Visual \cameraready{information} is thus essential for conveying musical elements like emotion and timing.
Consistent with prior work \cite{yoo2023understanding}, formative study participants reported using visual aids, including online videos. 
To address this, we designed ELMI to provide music videos with captions synchronized to the lyrics (\eg{},\circledigit{c} in \autoref{fig:system:glossing}), allowing users to fully grasp the nuances and accurately translate them.


\ipstart{DR2: Support discussion about nuances, performance, and translation of lyrics} 
\revised{Participants in formative study commonly engaged in discussions about song interpretation, mirroring how individuals use online forums to learn and share insights \cite{waldron2013youtube, o2019self, blom2014finding}.}
They also recognized the potential of tools like ChatGPT to support their workflow, aligning with recent work on using ChatGPT for sign language translation \cite{shahin2023chatgpt, othman2024comparative}.
Based on these findings, we incorporated an LLM-driven chat feature to ELMI, which allows discussions across four key topics: meaning, glossing, emoting, and timing. We intended to facilitate brainstorming and the exchange of ideas, allowing song-signers to gain different perspectives and supporting users in making artistic decisions.

\ipstart{DR3: Focus translation at a line level}
Unlike typical text, song lyrics are structured in lines rather than sentences and paragraphs, and they often feature shorter, repetitive phrasing \cite{choi2018computational}. 
We found from the formative study that song-signers typically gloss lyrics line by line. In response, we designed ELMI to facilitate gloss creation, discussion, and playback of the music video on a line-by-line basis.
This approach helps users break down the lyrics into manageable sections, allowing them to focus on translating each line, while also accommodating the iterative nature of refining individual lines to achieve the best possible interpretation. \revised{We initially considered supporting AI discussions about the general information of the song to help users familiarize themselves with it before translation. However, formative study participants rarely discussed such topics with ChatGPT, and most preferred static materials or performance videos for initial research. As a result, we limited AI discussions to the \cameraready{line-by-line translation.}}

\subsection{User Interface and Interaction Components}

ELMI operates in two modes: By default, the system is in the \textbf{Global Play} mode, where the user can play the entire video using a conventional video player (\circledigit{e} in \autoref{fig:teaser}). In this mode, the \textbf{Information Panel} (\autoref{fig:teaser}-\blackrectsmall{A}) serves as the user’s entry point for exploring the story and meaning of the song; it displays essential song details, such as genre, runtime, and behind-the-scenes insights, all sourced from the song description of Genius music database~\cite{genius}.
While the song is being played, the \textbf{Translation Panel} (\autoref{fig:teaser}-\blackrectsmall{B}) automatically scrolls to display the current line.

\begin{figure*}[]
    \centering
    \includegraphics[width=0.95\textwidth]{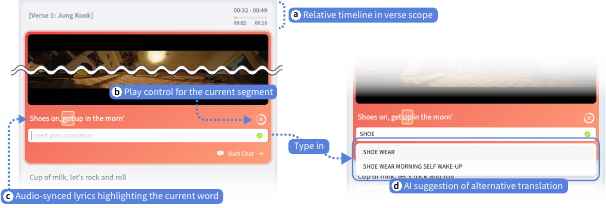}
    \caption{The lyric line translation pane. ELMI offers rich visual feedback to convey the song's timing; the user can check relative music position~\circledigit{a} at a verse level, as well as the line level~\circledigit{b}. While the music is being played, the corresponding lyric words highlight to enhance the user's sense of timing~\circledigit{c}. When the user is typing in the gloss, the system provides real-time suggestion of alternative translations in varied lengths~\circledigit{d}.}
    \Description{The figure depicts the lyric translation pane of the ELMI interface, which provides visual feedback to help users understand the timing of a song. The pane shows different levels of lyric display:
    (a) Verse-level music position is indicated to help the user follow along.
    (b) Line-level music position shows the current line of lyrics in context with the verse.
    (c) As the music plays, individual lyric words are highlighted to indicate precise timing.
    (d) When the user types gloss translations, the system suggests alternative translations in real time, offering various lengths and styles of glosses.}
    \label{fig:system:glossing}
\end{figure*}

The user can enter the \textbf{Line Loop} mode by selecting a lyric line on the Translation Panel. In this mode, the selected line is highlighted (\autoref{fig:teaser}-\blackrectsmall{B}, top), and the user can create and refine glosses  \revised{(\circledigit{d} in \autoref{fig:teaser})} with real-time feedback from the chatbot  \revised{(\circledigit{f} in \autoref{fig:teaser})}. 
In this mode, the Information Panel displays the AI-generated analysis results for the line \revised{with guidance on emotional expression such as non-manual signals (NMS)}, including a Mood section  \revised{(\circledigit{a} in \autoref{fig:teaser})} with hashtagged emotions, and a Performance Guide \revised{(\circledigit{b} in \autoref{fig:teaser})} offering recommendations on gestures and facial expressions (See \autoref{fig:teaser}-\blackrectsmall{A}). 

On the Translation Panel, the user can write glosses directly beneath each line of lyrics (See \autoref{fig:system:glossing}). 
\revised{As they type, the system provides real-time suggestions for one or two alternative versions of the current input gloss, which users can choose to adopt (See \circledigit{d} in \autoref{fig:system:glossing}).}
The user can still play and pause the video, but the current line segment is looped (See \circledigit{b} in \autoref{fig:system:glossing}).

\begin{figure*}[t]
    \centering
    \includegraphics[width=\textwidth]{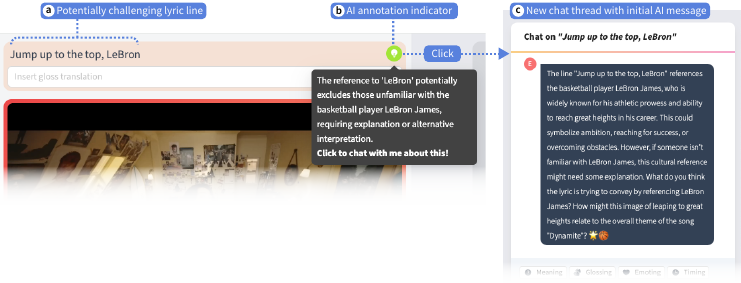}
    \caption{ELMI analyzes the lyrics in advance and marks noteworthy lines potentially challenging to translate~\circledigit{a}; In this case, an American basketball player, `LeBron' may not be recognized by users unfamiliar with the US sports scene. When the user hovers over the annotation indicator~\circledigit{b}, it shows a tooltip that invites the user to the discussion. If the user starts a chat thread by clicking the indicator, the AI will start the discussion directly~\circledigit{c}.}
    \Description{This screenshot figure demonstrates how ELMI analyzes lyrics and annotates lines that may be difficult to understand. The highlighted lyric line (a) is "Jump up to the top, LeBron," where 'LeBron' might be unfamiliar to users not well-versed in US sports. ELMI provides additional information when the user hovers over the AI annotation indicator (b). If the user starts a chat, a new thread opens with the AI proactively initiating the conversation (c).}
    \label{fig:system:proactive-chat}
\end{figure*}

ELMI also supports discussion with an LLM-driven chatbot on Chat Panel (\autoref{fig:teaser}-\blackrectsmall{C}) in Line Loop mode regarding artistic interpretation and other topics related to the selected line. The user can open a chat thread (\revised{\circledigit{f}} in \autoref{fig:teaser}) for each line by clicking the `Start Chat' button (See \autoref{fig:system:glossing}, left). Once a chat thread is open, users can either select from four shortcut buttons (Meaning, Glossing, Emoting, Timing; \revised{\circledigit{g}} in \autoref{fig:teaser}) or type their input directly into the chat box. In Global Mode, previous chat threads are displayed in a condensed format for fast navigation. To invite the user to the discussion on lines that are potentially challenging to translate, the system analyzes the lyrics in advance (\cf{},~\autoref{sec:system:conv}) and puts a light bulb icon (\circledigit{b} in \autoref{fig:system:proactive-chat}) on the noteworthy lines (\circledigit{a} in \autoref{fig:system:proactive-chat}). When the user initiates a chat thread for these lines, the chatbot directly starts a discussion on the meaning of the line (\circledigit{c} in \autoref{fig:system:proactive-chat}).

\subsection{Conversational Design: 4 Main Discussion Topics}
The dialogue interaction is structured around key discussion topics designed to support users throughout the translation process. 
\revised{The process begins with a preprocessing stage, where ELMI analyzes the lyrics to identify abstract or idiomatic language and generates a default gloss along with additional relevant data, such as emotional tones and timing cues. When a user requests glossing, ELMI suggests translations based on the default gloss, contextualized to align with the intended meaning of the lyrics.}
If a user provides their own gloss, the ELMI recognizes this input, updating its suggestions to align with the user’s input. 
This capability is managed through specialized chatbot modules that handle different discussion topics such as meaning extraction, gloss refinement, emoting, and timing: 
\begin{itemize}[leftmargin=*, itemsep=4pt, topsep=0pt]
    \item \meaningicon{}\textbf{Meaning} facilitates discussions about the message of the song, addressing poetic and ambiguous expressions, and considering variations due to different cultures, ages, and locations. 
    \item \glossingicon{}\textbf{Glossing} focuses on discussing sign selection, including non-manual signals (NMS). When there's no user input gloss, ELMI suggests ways to sign, while there's user input, ELMI suggests ways to improve signs. 
    \item \emotingicon{}\textbf{Emoting} centers on discussing the tone, vibe, and emotion of the song, encouraging conversations to explore ideas by using facial expression and body language. 
    \item \timingicon{}\textbf{Timing} addresses questions about adjusting the length of the gloss, offering users estimated timing without explicit numbers, multiple options for gloss length.
\end{itemize}


\subsection{Interacting with ELMI}

In this section, we illustrate how ELMI can be used through a scenario featuring Cindy, a Deaf song-signer with five years of experience. Cindy is preparing to interpret BTS's "Dynamite" in real-time at an upcoming concert and decides to use ELMI to create accurate glosses for the song.


After creating a new song project, Cindy enters the translation page. To familiarize herself with the song, she plays the music video, finding ELMI’s feature of highlighting sung words in real-time particularly helpful for understanding the song's timing.
Cindy begins her translation process by focusing on the first line of the song (\circledigit{\cameraready{d}} in \autoref{fig:teaser}). 
As she clicks on the line, the \textit{Information Panel} updates to provide a Mood and Performance guide (\blackrectsmall{A} in \autoref{fig:teaser}). She confirms that her feelings about the song align with ELMI’s interpretation, which suggests a joyful and uplifting mood (\circledigit{\cameraready{a}} \cameraready{in \autoref{fig:teaser}}). ELMI also offers guidance on \cameraready{non-manual signals (NMS) such as }facial expressions, advising Cindy to ``\texttt{raise your eyebrows and smile widely to emphasize the feeling of being in the stars.}'' (\circledigit{\cameraready{b}} \cameraready{in \autoref{fig:teaser}}).
Reflecting on the lyrics ``\textit{'Cause I, I, I'm in the stars tonight},'' Cindy writes her initial gloss as ``BECAUSE I PARTY TONIGHT.'' (\circledigit{\cameraready{d}} in \autoref{fig:teaser}.)

However, uncertain about the \meaningicon{}\textbf{meaning} of ``stars,'' she opens a new chat thread in the \textit{Chat Pane} and asks ELMI for clarification. ELMI suggests that ``stars'' could be a feeling of being extraordinary, celebrated, or having a bright moment (\circledigit{\cameraready{f}} \cameraready{in \autoref{fig:teaser}}). Taking this into account, Cindy revises her gloss to ``BECAUSE I SHINE, ENJOY  PARTY TONIGHT.'' She then synchronizes her gloss with the music, replaying the video in a loop. Noticing that her gloss is too long, Cindy asks ELMI for \timingicon{}\textbf{timing} advice and trims her gloss to ``I SHINE,  PARTY TONIGHT,'' confirming that it fits the song's timing.

When Cindy encounters the line ``Jump up to the top, LeBron,'' she’s unsure who ``LeBron'' refers to. She checks ELMI's tooltip (\circledigit{b} in \autoref{fig:system:proactive-chat}), which annotates that ``LeBron'' refers to the basketball player LeBron James, noting that this cultural reference may require additional context or finger spelling. Finding this line challenging to translate into ASL, Cindy clicks the \glossingicon{}\textbf{Glossing} button to get a suggestion from ELMI. ELMI suggests ``JUMP TOP F-S (FingerSpelling) `L-E-B-R-O-N','' but Cindy opts to use a classifier to depict a basketball action, so she types ``JUMP TOP CL-5 (basketball shooting).'' ELMI’s feedback confirms that her translation maintains the flow and meaning of the original line. Cindy then clicks \emotingicon{}\textbf{Emoting} and incorporates ELMI’s suggestions to express the energy of jumping through her performance.

Cindy continues to gloss the song by line, interacting with ELMI throughout the process. Once she finalizes her gloss, she uses ELMI to play the music video from the beginning, signing along with the song and using her glosses as a guide.

\begin{figure*}[b]
    \centering
    \includegraphics[width=\textwidth]{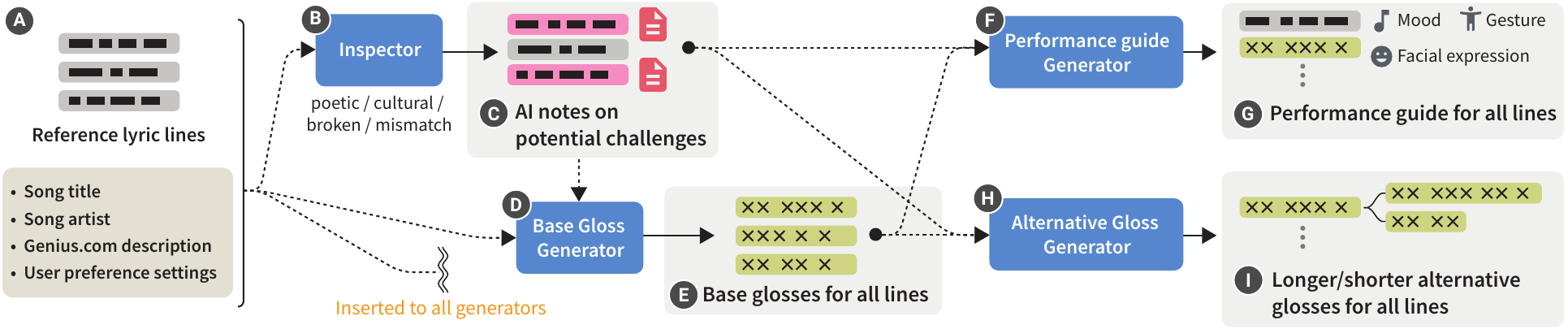}
    \caption{A pipeline for lyric analysis, which is part of pre-processing a song when the user creates a new project. Given the reference lyrics and metadata about the song and the user preference~\circledigit{A}, the pipeline chains four LLM inference modules (\circledigit{B}, \circledigit{D}, \circledigit{F}, and \circledigit{G}) to generate notes on potential challenges when translating specific lines \circledigit{C}, base gloss translation \circledigit{E}, performance guides for lines~\circledigit{G}, and the longer and shorter versions of each gloss line~\circledigit{I}.}
    \Description{This image depicts a pipeline for lyric analysis, which is part of the pre-processing step when a user creates a new project. The pipeline starts with the reference lyrics, song metadata, and user preferences (A). It then connects four LLM inference modules (B, D, F, G) to perform different tasks: identifying potential translation challenges for specific lines (C), generating a base gloss translation (E), providing performance guides for the lyric lines (G), and creating longer and shorter versions of each gloss line (I).}

    \label{fig:system:analysis}
\end{figure*}

\subsection{\revised{LLM-driven Conversational Pipelines}}\label{sec:system:conv}
The conversation system in ELMI leverages OpenAI's GPT-4 for preprocessing, intent classification, and chatbot interactions. The system uses tailored prompts specifically designed for each function to ensure accurate and contextually relevant responses. Refer to Appendix~\ref{appendix:prompt} for an example of the prompt used for the discussion with the chatbot.

\subsubsection{\revised{Lyric Analysis}}
When a user creates a new project, ELMI preprocesses the song by running the reference lyrics and metadata through four inference modules, as shown in the pipeline diagram (Fig. \ref{fig:system:analysis}).
The Line Inspector (\circledigit{B} in \autoref{fig:system:analysis}) annotates each line for challenges like poetic, cultural, or mismatched meanings. 
Based on these notes (\circledigit{C} in \autoref{fig:system:analysis}), the Base Gloss Generator (\circledigit{D} in \autoref{fig:system:analysis}) creates default glosses for all lines. Using these glosses (\circledigit{E} in \autoref{fig:system:analysis}), the Performance Guide Generator (\circledigit{F}) provides performance recommendations for gestures and expressions (\circledigit{G} in \autoref{fig:system:analysis}), while the Alternative Gloss Generator (\circledigit{H} in \autoref{fig:system:analysis}) produces alternative glosses and longer or shorter versions for all lines (\circledigit{I} in \autoref{fig:system:analysis}). \revised{The Performance Guide Generator also generates a new performance guide for the line when the user updates its translation.}

\subsubsection{\revised{Intent classifiers}}
User input is processed either through shortcut buttons (\circledigit{d} in \autoref{fig:teaser})) or direct input. 
When shortcut buttons are used, the system passes the intent directly. 
For open-ended queries, the intent classifier identifies the input and categorizes it into one of the four discussion topics.

\subsubsection{\revised{Chatbot}}
All user queries are saved, allowing the chatbot to remember previous conversations and use the dialogue history for future interactions. Based on the classified intent, the chatbot selects prompts in two ways. (1) If there is no user-generated gloss, it uses base inference results from the pre-processing stage. (2) If a user-generated gloss exists, the prompts are adapted to incorporate the gloss as input for further refinement.

ELMI's prompt input is designed to facilitate thoughtful interaction. As an active listener, ELMI encourages users to think critically, posing reflective questions about the lyrics and glosses. ELMI's role is to assist users in improving their sign performance, while its personality remains defined by clear communication, humor, empathy, and sensitivity. ELMI keeps the conversation focused on lyric interpretation while gently redirecting off-topic discussions and offering continuous support. Responses are designed to be concise, engaging, and thought-provoking.

\subsection{Implementation}
We developed ELMI's interface as a web application using TypeScript with React.js. The backend is powered by Python and built with the FastAPI framework, which serves as a REST API to manage both data and computational pipelines. These pipelines were implemented using LangChain to streamline multiple steps of LLM inferences, with data validation handled by Pydantic. A SQLite database is used to track the status of translations, messaging, and analysis results.



\revised{To provide fine-grained, real-time visual feedback for lyrics while playing audio, we used word-level timestamps (\ie, start and duration) of lyrics automatically generated through sequence matching and automatic speech recognition (ASR). The pipeline leverages reference lyrics from Genius.com~\cite{genius} lyric database and obtains word-level timestamps matching the lyrics with the ASR transcription results from OpenAI Whisper API~\cite{radford2022whisper}, which provides word-level timestamps for transcription. To improve accuracy, we first segmented the audio by lines using the subtitle information in the song's official YouTube video and generated word-level timestamps for each line. (Refer to Appendix~\ref{appendix:lyricalignment} where we describe the detailed pipeline steps.)}
\section{User study}
We conducted an exploratory user study with ELMI, employing observation methods to examine how a Large Language Model (LLM) chat agent can support song-signing gloss creation.
To ensure the effectiveness of the system and the study protocol, we refined both through two pilot sessions with hard-of-hearing song-signers. The study protocol was approved by our Institutional Review Board. 

\begin{table*}[t]
\sffamily
\small
	\def\arraystretch{1.3}\setlength{\tabcolsep}{0.25em}
		    \centering

\caption{Demographics, description, communication methods, and song-signing experience of our main study participants.}
\Description{This table presents the demographics, descriptions, communication methods, and song signing experience of the main study participants. The columns provide details on participant ID (PID), age and gender, a brief description of their hearing status, their role (e.g., Deaf artist, ASL interpreter, hobbyist), their experience with song signing (e.g., frequency and years of experience), and their preferred communication methods (e.g., ASL, PSE, talking, or signing).}
\label{tab:demographic}
\begin{tabular}{|l!{\color{gray}\vrule}c!{\color{tablegrayline}\vrule}m{0.16\textwidth}!{\color{tablegrayline}\vrule}m{0.18\textwidth}!{\color{tablegrayline}\vrule}l!{\color{tablegrayline}\vrule}m{0.22\textwidth}!{\color{tablegrayline}\vrule}m{0.10\textwidth}|}
\hline
\rowcolor{tableheader}
\textbf{PID} & \textbf{Age/Gender} & \textbf{Description} & \textbf{Role} & \textbf{Song signing} & \textbf{Communication} & \textbf{Language} \\ \hline
\textbf{\p{1}} & 37/M & Deaf\newline{}(Profound/hearing aids) & Hobbyist (Church worship) & Always/4 years & Sign or talk (not at the same time) & ASL/fluent \\ \arrayrulecolor{tablegrayline}\hline
\textbf{\p{2}} & 45/F & Deaf\newline{}(Profound/hearing aids) & Deaf artist & Sometimes/25 years & Sign or talk (not at the same time) & ASL/native \\ \hline
\textbf{\p{3}} & 48/F & Deaf\newline{}(Profound/hearing aids) & Official ASL interpreter & Often/30 years & Sign or talk (not at the same time) & ASL/native \\ \hline
\textbf{\p{4}} & 36/M & Hard of hearing\newline{}(Profound/hearing aids) & Deaf artist & Often/14 years & Sign or talk (not at the same time) & ASL/fluent \\ \arrayrulecolor{tablegrayline}\hline
\textbf{\p{5}} & 32/M & Hard of hearing\newline{}(mild/none) & Official ASL interpreter & Often/15 years & Sign or talk (not at the same time) & ASL/fluent \\ \hline
\textbf{\p{6}} & 43/F & Hard of hearing\newline{}(mild/none) & Hobbyist (Church worship) & Often/4 years & Talk only & PSE/moderate \\ \arrayrulecolor{black}\hline
\textbf{\p{7}} & 36/M & Hearing & Freelance ASL interpreter & Sometimes/4 years & Talk only & PSE/fluent \\ \arrayrulecolor{tablegrayline}\hline
\textbf{\p{8}} & 27/F & Hearing & Hobbyist & Sometimes/12 years & Talk only & PSE/moderate \\ \hline
\textbf{\p{9}} & 25/M & Hearing & Deaf school teacher & Sometimes/4 years & Sign or talk (not at the same time & ASL/moderate \\ \hline
\textbf{\p{10}} & 36/Non-binary & Hearing & ASL artist & Always/12 years & Sign and talk at the same time & ASL/fluent \\ \hline
\textbf{\p{11}} & 67/F & Hearing & Hobbyist (Church worship) & Often/20 years & Sign or talk (not at the same time) & ASL/moderate \\ \hline
\textbf{\p{12}} & 36/M & Hearing & Hobbyist & Rarely/7 years & Talk only & ASL/novice \\ \hline
\textbf{\p{13}} & 30/F & Hearing & Official ASL interpreter & Rarely/8 years & Sign or talk (not at the same time) & ASL/fluent \\ \arrayrulecolor{black}\hline
\end{tabular}
\end{table*}

\subsection{Participants}
We recruited \textit{13} song-signers (P1--13) through various channels, including social media platforms such as Facebook groups and Reddit, email lists, word of mouth, and snowball sampling. 
\revised{Three participants self-identified as deaf, three as hard of hearing, and the remaining seven as hearing.}
Our inclusion criteria required participants to (1) have completed at least three song-signing projects before the study, (2) be a minimum of 18 years old, (3) be comfortable reading and writing in English, and (4) use American Sign Language (ASL) or Pidgin Signed English (PSE) in their song-signing work. \autoref{tab:demographic} summarizes the demographic of our study participants. 
Participants were aged between 25 and 67 years ($M = 38.3$) and included 9 males, 4 females, and 1 non-binary individual.
Ten out of 13 participants communicated primarily in ASL, with others using Pidgin Signed English (PSE). 
On average, participants had 12.2 years of experience in song-signing, with their involvement ranging from 4 to 30 years, driven by professional and personal motivations, ranging from ASL interpreters and artists to hobbyists and Deaf school teachers.

\subsection{Songs for Translation}
\cameraready{Our study involved translating two songs.} First, participants translated the same song (referred to as the \textbf{\assignedsong{}} hereinafter) so that we could observe common interaction patterns and diversity of translation. 
Second, each participant chose one song (referred to as the \textbf{\selectedsong{}} hereinafter) they wanted to translate.


For the \assignedsong{}, we selected ``Butter'' by ``BTS'' \footnote{\url{https://www.youtube.com/watch?v=WMweEpGlu_U}}, considering its moderate complexity, clear emotional tone, and being free of explicit content. The song conveys themes of self-discovery, resilience, and love~\cite{yuliana2024bts, sitompul2022translation}.
The song's lyrics, which include multiple types of figurative language such as metaphors, similes, hyperbole, and personification \cite{gradistina2023figurative, fauqi2020metaphorical, aryawan2022deixis}, which was suitable to test ELMI's ability to interpret and gloss complex expressions.

For the \selectedsong{}, we asked participants to select up to three candidate songs that meet the following criteria: (1) written in English, (2) with an official music video available on YouTube, and (3) with reference lyrics available online. 
Participants were encouraged to choose preferred songs, as this would likely lead to meaningful interactions. \cameraready{The list of selected songs and YouTube videos used in the study is provided in Appendix \ref{appendix:selectedsongs}.}

\subsection{Study Procedure}

\subsubsection{Pre-study Preparation}
Before the main study session, participants completed a pre-study survey, which collected demographic information, hearing description, song-signing experience, ASL expertise, and details about the tools and methods they used for song-signing and gloss creation. Participants also submitted three candidates for the \selectedsong{}.

\subsubsection{Main Study Session}
Each participant engaged in a 2-hour main study session remotely on Zoom. Depending on their hearing capability, the session was carried on via spoken language with live captioning, Zoom chats, or with an ASL interpreter. The Zoom session was audio- and video-recorded for analysis.

\ipstart{Briefing and Tutorial}
We first described the goal of our study and the study protocol, covering the overall process of creating a project, glossing, and chatting. As a tutorial for the ELMI interface, participants shared their screen and tried ELMI with the example song, ``Viva La Vida'' by Coldplay. Participants were allowed to practice with the tool until they felt confident about the interface. This phase took about 20 minutes.

\ipstart{Song Translation}
Participants created glosses using ELMI for two projects, translating \assignedsong{} and \selectedsong{}. Due to time constraints, we asked participants to translate only the first verse of each song. We allowed up to 30 minutes for each song but also asked not to rush to finish within the time limit and to prioritize the translation quality as usual.

\ipstart{Debriefing}
We conducted a semi-structured interview around 30 minutes at the end of the session. We asked participants about their experience with ELMI, the quality of translation they created, the benefits and concerns of ELMI, and follow-up questions based on our observations in the translation phase. We also asked them how they envision ELMI to be applied to their workflow and their behavioral intention to use the tool after the study. We offered participants a 75 USD Amazon gift card for their participation.

\subsubsection{Post-study Assignments}
Due to the time constraints, most participants did not finish the two songs during the session. Hence, we asked participants to finish verse 1 of the \assignedsong{} after the study session, offering additional incentives of 25 USD. Participants also optionally completed the \selectedsong{} and submitted a post-study survey, where we asked about their experience of using ELMI in their own environment. This survey also included a 5-point Likert scale questionnaire consisting of questions derived from the Technology Acceptance Model (TAM)~\cite{davis1989technology} regarding the self-assessed quality of translation, ease of use, and willingness to use ELMI. Participants who finished this optional task were compensated with an additional 80 USD. All participants except \p{5} and \p{13} completed post-study tasks.



\subsection{Analysis}
\ipstart{Glossing}
We analyzed the translation results of the \assignedsong{} from 11 participants who completed the first verse. 
Key metrics such as the number of lines completed and the average word count were analyzed using Pandas\footnote{\url{https://pandas.pydata.org/}} in Python.

\begin{sloppypar}
\ipstart{Chats}
We examined chat threads of both \assignedsong{} and \selectedsong{} from 11 participants who completed post-study assignments. 
We defined a \textit{turn} as a single message exchange, with \textit{user turns} representing participant messages and \textit{system turns} representing those by ELMI. Using the Pandas Python package, we computed various descriptive statistics, such as the total number of turns per session. Additionally, we categorized conversation turns using a top-down approach based on four discussion topics, and researchers independently coded the dialogues of three participants.\end{sloppypar}

\ipstart{Interviews and Surveys}
We also analyzed debriefing transcripts, post-study surveys, and video recordings and screen captures, to understand participants' perspectives and expectations regarding ELMI. 
Thematic analysis~\cite{braun2012thematic} was employed to code and examine the transcripts. \revised{Initial coding was performed by two researchers independently, followed by iterative discussions to resolve discrepancies and refine the codes. Recurring codes were then grouped into broader categories and further distilled into key themes through team discussions. } 
Three key themes emerged: \revised{interactions with ELMI, including the four discussion topics; perceptions of ELMI's tone and its influence on translation ownership; and the benefits and challenges experienced while using ELMI.}
\section{Results}
This section presents the study’s findings across five key areas: (1) Translation Outputs, (2) Conversations with ELMI, (3) Perceptions of ELMI and (4) Strengths and Limitations of Using ELMI.

\subsection{Translation Outcome}

\def\lyriccolumnwidth{0.20\textwidth}
\def\examplecolumnwidth{0.40\textwidth}

\begin{table*}[b]
\sffamily
\small
	\def\arraystretch{1.1}\setlength{\tabcolsep}{0.25em}
		    \centering
\caption{Summary of participants' translation of \revised{the first verse of} \assignedsong{}, including the word count of gloss lines and the average of pairwise overlap coefficient across participants' gloss words for the same line. 0\% of average overlap means that all glosses from participants include distinct words.}
\Description{This table summarizes participants' translations of the first verse (9 lines) of the song "BTS-Butter," showing the word count of gloss lines and the average pairwise overlap coefficient across participants' gloss words for each lyric line. The table includes columns for each lyric line, the original English lyric, the language (ASL or PSE), an example gloss from participants, the minimum, maximum, mean, and standard deviation of the word count in glosses, and the average overlap percentage. The overlap percentage indicates how much participants' glosses for the same line use similar words. Each row provides detailed data for specific lyric lines.}
\label{tab:results:translation}
\begin{tabular}{|c!{\color{gray}\vrule}m{\lyriccolumnwidth}!{\color{gray}\vrule}c!{\color{lightgray}\vrule}m{\examplecolumnwidth}!{\color{lightgray}\vrule}cccc!{\color{lightgray}\vrule}c|}
\hline
\rowcolor[HTML]{EFEFEF} 
\cellcolor[HTML]{EFEFEF}                                & \cellcolor[HTML]{EFEFEF}                                 & \cellcolor[HTML]{EFEFEF}                                & \cellcolor[HTML]{EFEFEF}                                         & \multicolumn{4}{c!{\color{lightgray}\vrule}}{\cellcolor[HTML]{EFEFEF}\textbf{Gloss word count}}                                                                                  & \cellcolor[HTML]{EFEFEF}                                            \\
\rowcolor[HTML]{EFEFEF} 
\multirow{-2}{*}{\cellcolor[HTML]{EFEFEF}\textbf{Line}} & \multirow{-2}{*}{\cellcolor[HTML]{EFEFEF}\textbf{Lyric}} & \multirow{-2}{*}{\cellcolor[HTML]{EFEFEF}\textbf{Lang}} & \multirow{-2}{*}{\cellcolor[HTML]{EFEFEF}\textbf{Example gloss}} & {\color[HTML]{333333} \textbf{Min}} & {\color[HTML]{333333} \textbf{Max}} & {\color[HTML]{333333} \textbf{Mean}} & {\color[HTML]{333333} \textbf{SD}} & \multirow{-2}{*}{\cellcolor[HTML]{EFEFEF}\parbox[t]{0.10\textwidth}{\centering \textbf{Avg. overlap}}} \\
\hline
                                                        &                                                          & ASL                                                     &                                                   ME SAME-AS BUTTER SMOOTH (\p{10})               & 2                                   & 5                                   & 3.50                                 & 1.22                                & 67.33\%                                                              \\\arrayrulecolor{tablegrayline}\cline{3-9}
\multirow{-2}{*}{1}                                     & \multirow{-2}{*}{Smooth like butter}                     & PSE                                                     &                      SMOOTH LIKE BUTTER (\p{6}, \p{8}, \p{11})                                            & 3                                   & 5                                   & 3.80                                 & 1.10                                & 62.00\%                                                              \\\arrayrulecolor{lightgray}\hline
                                                        &                                                          & ASL                                                     &                                       PERSON DANGEROUS DISGUISE (\p{3})                           & 2                                   & 3                                   & 2.83                                 & 0.41                                & 0.00\%                                                               \\\arrayrulecolor{tablegrayline}\cline{3-9}
\multirow{-2}{*}{2}                                     & \multirow{-2}{*}{\parbox[t]{\lyriccolumnwidth}{Like a criminal\newline{}undercover}}             & PSE                                                     &                   LIKE SECRET THIEF (\p{7})                                               & 3                                   & 5                                   & 3.60                                 & 0.89                                & 32.50\%                                                             \\\arrayrulecolor{lightgray}\hline
                                                        &                                                          & ASL                                                     &                 GUN [LCL``shoot''] (\p{2})                                                 & 1                                   & 4                                   & 2.83                                 & 1.17                                & 11.11\%                                                              \\\arrayrulecolor{tablegrayline}\cline{3-9}
\multirow{-2}{*}{3}                                     & \multirow{-2}{*}{Gon' pop like trouble}                  & PSE                                                     &                  APPEAR LIKE TROUBLE (\p{8}, \p{11})                                                & 3                                   & 4                                   & 3.20                                 & 0.45                                & 50.00\%                                                              \\\arrayrulecolor{lightgray}\hline
                                                        &                                                          & ASL                                                     &               PENETRATE HEART HEART [CL ``pump''] (\p{10})                                                   & 2                                   & 5                                   & 3.17                                 & 1.17                                & 53.33\%                                                             \\\arrayrulecolor{tablegrayline}\cline{3-9}
\multirow{-2}{*}{4}                                     & \multirow{-2}{*}{\parbox[t]{\lyriccolumnwidth}{Breakin' into your heart\newline{}like that}}     & PSE                                                     &                                         ME ENTER YOUR HEART (\p{6})                         & 3                                   & 5                                   & 4.40                                 & 0.89                                & 22.33\%                                                             \\\arrayrulecolor{lightgray}\hline
                                                        &                                                          & ASL                                                     &            [CL:1 ``person wearing sunglasses, looking cool''] [HEAD-tilt ``same as music video''] (\p{2})                                                      & 0                                   & 4                                   & 2.83                                 & 1.17                                & 22.22\%                                                             \\\arrayrulecolor{tablegrayline}\cline{3-9}
\multirow{-2}{*}{5}                                     & \multirow{-2}{*}{Cool shade stunner}                     & PSE                                                     &               COOL SHADE AWESOME PERSON (\p{7})                                                   & 1                                   & 4                                   & 3.00                                 & 1.22                                & 9.17\%                                                              \\\arrayrulecolor{lightgray}\hline
                                                        &                                                          & ASL                                                     &               [HEAD-nod] THANK MOTHER (\p{2})                                                   & 2                                   & 6                                   & 3.50                                 & 1.38                                & 21.67\%                                                              \\\arrayrulecolor{tablegrayline}\cline{3-9}
\multirow{-2}{*}{6}                                     & \multirow{-2}{*}{\parbox[t]{\lyriccolumnwidth}{Yeah, I owe it all\newline{}to my mother}}        & PSE                                                     &                                     ME OWE MOM EVERYTHING (\p{8})                             & 3                                   & 7                                   & 5.40                                 & 1.82                                & 27.86\%                                                             \\\arrayrulecolor{lightgray}\hline
                                                        &                                                          & ASL                                                     &                 SUNSHINE [CL:5 ``coming from me''] (\p{10})                                                 & 1                                   & 4                                   & 2.67                                 & 1.03                                & 55.56\%                                                              \\\arrayrulecolor{tablegrayline}\cline{3-9}
\multirow{-2}{*}{7}                                     & \multirow{-2}{*}{Hot like summer}                        & PSE                                                     &               HOT LIKE SUMMER (\p{6}, \p{8})                                                   & 3                                   & 5                                   & 4.00                                 & 1.00                                & 79.83\%                                                             \\\arrayrulecolor{lightgray}\hline
                                                        &                                                          & ASL                                                     &                    YOU SWEAT YES (\p{12})                                              & 2                                   & 5                                   & 3.67                                 & 1.21                                & 46.11\%                                                             \\\arrayrulecolor{tablegrayline}\cline{3-9}
\multirow{-2}{*}{8}                                     & \multirow{-2}{*}{\parbox[t]{\lyriccolumnwidth}{Yeah, I'm makin' you\newline{}sweat like that}}   & PSE                                                     &                                     ME CAUSE YOU SWEAT (\p{4}, \p{6}, \p{8})                             & 2                                   & 8                                   & 4.60                                 & 2.79                                & 40.36\%                                                             \\\arrayrulecolor{lightgray}\hline
                                                        &                                                          & ASL                                                     &                    RELAX ENJOY (\p{2})                                              & 1                                   & 4                                   & 2.00                                 & 1.10                                & 6.67\%                                                              \\\arrayrulecolor{tablegrayline}\cline{3-9}
\multirow{-2}{*}{9}                                     & \multirow{-2}{*}{Break it down}                          & PSE                                                     &                  DANCE (\p{8})                                                & 1                                   & 3                                   & 1.80                                 & 0.84                                & 0.00\%                                                               \\
\arrayrulecolor{black}\hline
\end{tabular}
\end{table*}

We present the glossing results for \assignedsong{}, BTS's ``Butter'' to examine how ELMI affected \revised{the diversity of translations produced by participants for the same lyrics.} We analyzed the first three parts (Verse 1, Pre-chorus, Chorus), comprising 19 lines with a total of 105 words. Among the participants, six (3 d/Deaf, 3 hearing) used ASL, while five (2 d/Deaf, 3 hearing) used PSE.
Participants spent varied time translating lyrics during the main session, 
completing between 3\delete{ [\p{10}]} and 19\delete{ [\p{7}, \p{8}]} lines, with an average completion of 9.82 lines ($SD = 5.25$, $min = 2$\delete{ [\p{1}]}, $max = 19$ \delete{[\p{10}]}).

\autoref{tab:results:translation} summarizes the \revised{translation result of the first verse of} \assignedsong{} by 11 participants who completed the post-study assignments. 
Participants produced a wide variety of gloss translations, incorporating non-manual signals (NMS), such as classifiers, visual vernacular, facial expressions, and body language.
On average, participants used 3.72 signs per line ($SD = 1.38$, $min = 1.9$\delete{ [line 9]}, $max = 5.72$\delete{ [line 15]}), with 0.44 NMS per line ($SD = 0.91$, $min = 0.09$\delete{ [line 3, 13]}, $max = 1.09$\delete{ [line 14]})\revised{. There was a high individual variance in both the number of signs and NMS.}
\revised{We did not observe a significant difference between the d/Deaf and hearing participants (3.61 signs and 0.75 NMS per line for d/Deaf, and 3.81 signs and 0.18 NMS per line for hearing participants), although d/Deaf participants tended to use NMS more actively.}

To assess the diversity of translation among participants, we calculated the average overlap coefficient\footnote{Calculated with the length of intersection of the word sets A and B divided by the length of min(A, B)} of pairwise combinations of manual signs for each lyric line (See \autoref{tab:results:translation}, `Average Overlap'). The average overlap coefficient of ASL signs of 19 lines was 36.36\% ($SD=20.53\%$, $min=0\%$\delete{ [line 2]}, $max=72.22\%$\delete{ [line 10]}), and that of PSE was 38.39\% ($SD=21.62\%$, $min=0\%$\delete{ [line 9]}, $max=80.38\%$\delete{ [line 10]}), suggesting that participants shared around one-third of signs for the same line. 
For instance, when translating line 2, ``\textit{Like a criminal undercover},'' PSE participants seemed to assign \cameraready{literal} signs such as ``LIKE SECRET THIEF'' [\p{7}]. In contrast, ASL participants produced more diverse and distinct signs such as ``ITSELF THEFT OVERLOOK'' [\p{1}] or ``PERSON DANGEROUS DISGUISE'' [\p{3}].

\subsection{Conversation with ELMI}

\begin{table}[]
\sffamily
\small
	\def\arraystretch{1.25}\setlength{\tabcolsep}{0.4em}
		    \centering
    \caption{Average number of messages per thread and the breakdowns of user-generated messages from the shortcut buttons (shortcut messages) and those typed manually (manual messages) per participant.}
    \Description{This table compares various metrics related to message activity for two different songs, "assigned song" and "selected song." The metrics include the average number of messages per thread, average number of shortcut messages, and average number of manual messages, with breakdowns for d/Deaf participants and hearing participants. For each metric, the table presents the mean and standard deviation (SD) for both d/Deaf and hearing participants across both songs.}
    \label{tab:chat}
    \begin{tabular}{|m{0.38\columnwidth}!{\color{lightgray}\vrule}rr!{\color{lightgray}\vrule}rr|}
    \hline
    \rowcolor{tableheader} & \multicolumn{2}{c}{\textbf{\assignedsong{}}} & \multicolumn{2}{!{\color{lightgray}\vrule}c|}{\textbf{\selectedsong{}}} \\ \hline
    \rowcolor{tableheaderlight}\textbf{\# of messages per thread} & 5.91 & ($SD$ = 1.97) & 5.14 & ($SD$ = 1.61) \\ \arrayrulecolor{tablegrayline}\hline
    \deafwrap{d/Deaf} only & 5.75 & ($SD$ = 1.72)  & 4.74 & ($SD$ = 1.98)\\ \hline
    \hearingwrap{hearing} only & 6.03 & ($SD$ = 2.31) & 5.47 & ($SD$ = 1.34)\\ \arrayrulecolor{black}\hline
    \rowcolor{tableheaderlight}\textbf{\# of shortcut messages} & 13.36 & ($SD$ = 11.93)& 11.55 & ($SD$ = 9.77)\\ \arrayrulecolor{tablegrayline}\hline
    \deafwrap{d/Deaf} only & 7.20 & ($SD$ = 8.50)& 5.40 & ($SD$ = 7.09)\\ \hline
    \hearingwrap{hearing} only & 18.5 & ($SD$ = 12.55)& 16.67 & ($SD$ = 9.03)\\ \arrayrulecolor{black}\hline
    \rowcolor{tableheaderlight}\textbf{\# of manual messages} & 14.64 & ($SD$ = 17.98)& 11.64 & ($SD$ = 12.71)\\ \arrayrulecolor{tablegrayline}\hline
    \deafwrap{d/Deaf} only & 16.2 & ($SD$ = 19.05)& 8.40 & ($SD$ = 12.58)\\ \hline
    \hearingwrap{hearing} only & 13.33 & ($SD$ = 18.76)& 14.33 & ($SD$ = 13.32)\\ \arrayrulecolor{black}\hline
    \end{tabular}%
\end{table}

\begin{table}[]
\sffamily
\small
	\def\arraystretch{1.25}\setlength{\tabcolsep}{0.4em}
		    \centering
    \caption{Average number of messages per participant by discussion topic.}
\Description{This table provides an analysis of chat discussion topics, comparing the number of messages exchanged for various topics across two songs, "assigned song" and "selected song." The topics include Meaning, Glossing, Emoting, and Timing. For each topic, the table shows the average number of messages exchanged, along with the standard deviation (SD), broken down by d/Deaf participants and hearing participants.}
\label{tab:feature}
    \begin{tabular}{|m{0.38\columnwidth}!{\color{lightgray}\vrule}rr!{\color{lightgray}\vrule}rr|}
    \hline
    \rowcolor{tableheader} & \multicolumn{2}{c}{\textbf{\assignedsong{}}} & \multicolumn{2}{!{\color{lightgray}\vrule}c|}{\textbf{\selectedsong{}}} \\ \hline
    \rowcolor{tableheaderlight}\textbf{\# of Meaning messages} & 7.90 & ($SD$ = 6.25)& 4.55 & ($SD$ = 4.10)\\ \arrayrulecolor{tablegrayline}\hline
    \deafwrap{d/Deaf} only & 9.20 & ($SD$ = 6.87)& 5.00 & ($SD$ = 5.00)\\ \hline
    \hearingwrap{hearing} only & 6.83 & ($SD$ = 6.11)& 4.17 & ($SD$ = 3.66)\\ \arrayrulecolor{black}\hline
    \rowcolor{tableheaderlight}\textbf{\# of Glossing messages} & 14.64 & ($SD$ = 13.98)& 11.90 & ($SD$ = 11.28)\\ \arrayrulecolor{tablegrayline}\hline
    \deafwrap{d/Deaf} only & 10.00 & ($SD$ = 7.78)& 5.00 & ($SD$ = 5.00)\\ \hline
    \hearingwrap{hearing} only & 18.50 & ($SD$ = 6.11)& 17.67 & ($SD$ = 12.13)\\ \arrayrulecolor{black}\hline
    \rowcolor{tableheaderlight}\textbf{\# of Emoting messages} & 4.10 & ($SD$ = 4.57)& 4.63 & ($SD$ = 8.20)\\ \arrayrulecolor{tablegrayline}\hline
    \deafwrap{d/Deaf} only & 3.00 & ($SD$ = 3.08)& 1.80 & ($SD$ = 2.49)\\ \hline
    \hearingwrap{hearing} only & 5.00 & ($SD$ = 5.66)& 7.00 & ($SD$ = 10.71)\\ \arrayrulecolor{black}\hline
    \rowcolor{tableheaderlight}\textbf{\# of Timing messages} & 1.36 & ($SD$ = 1.12)& 2.10 & ($SD$ = 2.07)\\ \arrayrulecolor{tablegrayline}\hline
    \deafwrap{d/Deaf} only & 1.20 & ($SD$ = 1.30)& 2.00 & ($SD$ = 2.55)\\ \hline
    \hearingwrap{hearing} only & 1.50 & ($SD$ = 1.05)& 2.17 & ($SD$ = 1.35)\\ \arrayrulecolor{black}\hline
    \end{tabular}%
\end{table}

\begin{figure}[b]
    \centering
    \includegraphics[width=\linewidth]{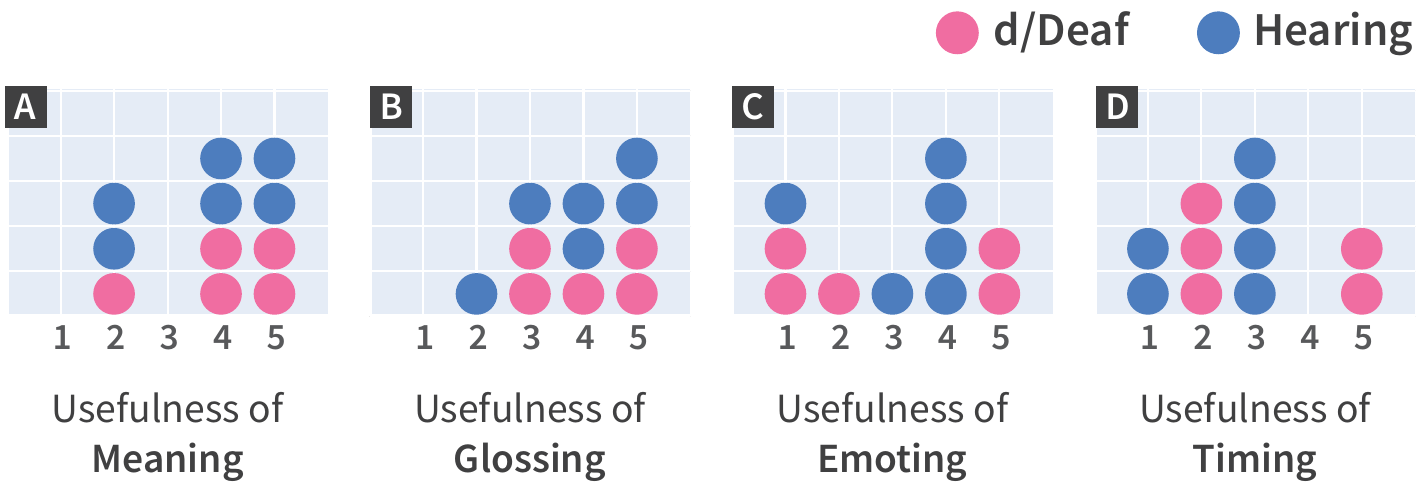}
    \caption{Likert-scale ratings of the 4 discussion topics (1: not useful 5: extremely useful).}
   \Description{This figure presents a histogram of Likert-scale ratings on the usefulness of the four discussion topics—Meaning, Glossing, Emoting, and Timing—based on a 1 to 5 scale, where 1 indicates "not useful" and 5 indicates "extremely useful." The ratings are visually represented with pink circles for d/Deaf participants and blue circles for hearing participants.}
    \label{fig:rating:chat}
\end{figure}

\begin{sloppypar}
Overall, participants opened a total of 222 threads, with 114 threads from \assignedsong{} and 108 from \selectedsong{}. These threads comprised 1,346 messages—730 from \assignedsong{} and 616 from \selectedsong{}, featuring 563 user messages (308 from \assignedsong{}, 255 from \selectedsong{}) and 775 system messages (414 from \assignedsong{}, 361 from \selectedsong{}).
For \assignedsong{}, participants opened an average of 10.36 threads from 19 lines ($SD = 5.32$, $min = 3$\delete{ [\p{10}]}, $max = 17$\delete{ [\p{7}, \p{8}]}), with each thread containing 5.91 messages ($SD = 1.97$, $min = 3.5$\delete{ [\p{4}]}, $max = 10.53$\delete{ [\p{7}]}).
As for \selectedsong{}, participants opened an average of 9.82 threads from 47.36 lines ($SD = 7.07$, $min = 1$\delete{ [\p{1}]}, $max = 23$\delete{ [\p{9}]}), with each thread containing 5.14 messages ($SD = 1.61$, $min = 3$\delete{ [\p{1}, \p{4}]}, $max = 7.73$\delete{ [\p{3}]}) (See \autoref{tab:chat}).
To interact with ELMI, participants either clicked on one of four shortcut buttons (See \circledigit{d} in \autoref{fig:teaser}) or typed manual inputs. Participants used both shortcut message and manual message in a similar manner (See \autoref{tab:chat}). 
\revised{Regarding the four discussion topics, participants rated them as both useful (See \autoref{fig:rating:chat}) and frequently used (See \autoref{tab:feature}) in the following order: \glossingicon{}glossing, \meaningicon{}meaning, \emotingicon{}emoting, and \timingicon{}timing.}\end{sloppypar}

\bpstart{\meaningicon{}Meaning} was frequently discussed at the beginning of their threads to explore how to interpret specific words, 7.9 times ($SD = 6.25$, $min = 0$, $max = 17$) for \assignedsong{} and 4.55 times ($SD = 4.1$, $min = 0$, $max = 12$ for \selectedsong{} (See \autoref{tab:feature}).
\meaningicon{}Meaning was rated 3.81 on average ($SD = 1.25$, $min = 2$, $max = 5$), where d/Deaf participants gave 4 ($SD = 1.22$), and hearing \revised{participants} gave 3.66 ($SD = 1.37$)  (See \autoref{fig:rating:chat}).
\p{6} engaged with ELMI to interpret ``Cool shade stunner''. ELMI broke down the metaphor and encouraged \p{6} to consider the symbolic meaning and how it could be visually represented (See \autoref{dial:meaning}).

\begin{dialogue}[h]
\begin{quotetable}
\quotebotline{The phrase ``Cool shade stunner'' is poetic, employing a metaphor to convey a style or attitude that may be difficult to interpret directly into sign language.
\leavevmode\newline{}\null\newline{}
What do you think the author means by ``cool shade''?☘
\leavevmode\newline{}Is it about being physically cool or could it symbolize something else, like a relaxed, confident style?✌
\leavevmode\newline{} Is there a ``cool shade'' sign that might capture that feeling?}
\quoteuserline{\p{6}}{I think it's symbolic of someone being really suave and smooth to get into someone's heart.}
\quotebotline{That's an interesting perspective, [P6's nickname]! ❣  The idea of someone being suave and smooth certainly fits the essence of the lyric. Now, thinking about translating that into PSE, how would you visually express ``suave'' and ``smooth''? ✨ Do you think you'd use certain movements or expressions to convey that relaxed confidence we talked about?}
\end{quotetable}
\caption{Example conversation on \meaningicon{}Meaning}
\Description{This dialogue presents an example conversation on the topic of Meaning. The chatbot initiates the discussion by analyzing the phrase ``Cool shade stunner'' from the song lyrics, explaining that it is poetic and metaphorical, which might pose challenges for sign language translation. The chatbot then asks the user, identified as participant P6, what they think ``cool shade'' symbolizes and whether a sign exists to convey that meaning. P6 responds by interpreting ``cool shade'' as symbolic of someone being suave and smooth in winning someone's heart. The chatbot follows up by asking P6 to consider how they would visually express ``suave'' and ``smooth'' in PSE, with emphasis on conveying relaxed confidence through movements and expressions.}
\label{dial:meaning}
\end{dialogue}

Participants valued its ability to clarify and contextualize song lyrics, enabling them move beyond literal translations to more nuanced, expressive interpretations. 
As \p{2} noted, ``\textit{In the past, my translations were usually very literal. I didn’t think about the abstract meaning. But now, with ELMI, I realize I can create more visual interpretations. It’s not just a word-for-word translation but something more engaging.}''
The discussion was particularly effective in interpreting poetic language, offering alternative perspectives that helped participants explore different interpretations.

\bpstart{\glossingicon{}Glossing} was the most frequently used and highly rated topic: 14.64 times ($SD = 13.98$, $min = 1$, $max = 50$) for \assignedsong{} and 11.9 times ($SD = 11.28$, $min = 0$, $max = 38$) for \selectedsong{} (See \autoref{tab:feature}).
It had an average rating 3.9 ($SD = 1.04$, $min = 2$, $max = 5$), 4  from d/Deaf \revised{participants} ($SD = 1$), and 3.83 from hearing \revised{participants} ($SD = 1.17$) (See \autoref{fig:rating:chat}).
Song-signers often relied on it to get feedback on their glosses.
For example, \p{7} asked ELMI for guidance on a specific signing selection, and ELMI suggested several ways to combine signs to enhance expressive quality (See \autoref{dial:glossing}). 

\begin{dialogue}[h]
\begin{quotetable}
\quoteuserline{\p{7}}{Using the ``chilled girl'' sign along with the ``butter'' sign is a good match?}
\quotebotline{Using ``CHILLED GIRL'' with the ``BUTTER'' sign could indeed be a creative match, as it blends the relaxed and smooth concepts together in a visually expressive way.
\newline{}\null\newline{}
Here’s how you might combine them: \newline{}
1.\textbf{``CHILLED''} - Sign ``RELAX'' or ``COOL'' to capture the essence of being chilled.\newline{}
2. \textbf{``GIRL''} - Sign ``GIRL'' with a thumb brushing the cheek.\newline{}
3. \textbf{``BUTTER''} - Add the sign for ``BUTTER,'' which you can do by mimicking a smooth, spreading motion with your hand, like spreading butter on bread. 
\newline{}
So your combined gloss might look like this: \textbf{``RELAX GIRL BUTTER''} or \textbf{``COOL GIRL BUTTER''} with smooth, flowing motions to emphasize the cool and effortless vibe. ❄ ♨
\newline{}\null\newline{}
Does this combination feel like it captures the meaning and essence you're going for?\newline{}
Would you like to explore any additional elements to enhance it? 
}
\end{quotetable}
\caption{Example conversation on \glossingicon{}Glossing.}
\Description{This dialogue illustrates an example conversation on the topic of glossing. Participant P7 asks whether using the sign for ``chilled girl'' along with the sign for ``butter'' is a good match. The chatbot responds affirmatively, explaining that this combination creatively captures the relaxed and smooth concepts of the lyrics. The bot breaks down the signs: ``CHILLED'' could be signed as ``RELAX'' or ``COOL,'' ``GIRL'' is signed by brushing the thumb against the cheek, and ``BUTTER'' involves a smooth, spreading motion with the hand. The chatbot suggests a possible gloss as ``RELAX GIRL BUTTER'' or ``COOL GIRL BUTTER'' and encourages P7 to use smooth, flowing motions to emphasize the cool and effortless vibe. The bot also invites P7 to explore additional elements to enhance the gloss.}
\label{dial:glossing}
\end{dialogue}

Both d/Deaf and hearing participants found the \glossingicon{}Glossing valuable for exploring alternative options for specific phrases or signs, enabling creativity and adjustments based on the context and tone of the song. 
The feedback provided by ELMI was also well-received, as \p{9} noted, ``\textit{Glossing is more effective when I suggested my own ideas, as ELMI explains synonymous words and also offers suggestions on how to convey the message.}''

\bpstart{\emotingicon{}Emoting} was used 4.1 times ($SD = 4.57$, $min = 0$, $max = 14$) for \assignedsong{} and 4.63 times ($SD = 8.2$, $min = 0$, $max = 28$) for \selectedsong{} (See \autoref{tab:feature}). 
\emotingicon{}Emoting  had an average rating  3.09 ($SD = 1.58$, $min = 1$, $max = 5$), where d/Deaf participants gave 2.8 ($SD = 2.04$), and hearing \revised{participants} gave 3.33 ($SD = 1.21$)  (See \autoref{fig:rating:chat}).
It provided participants with suggestions on how to enhance the performance of specific lyrics through facial expressions and body language. For example, \p{3} asked about using ``raised eyebrows'' after finalizing her gloss ``SUDDENLY DANGER'' to translate the phrase ``Gon' pop like trouble.'' (See \autoref{dial:emoting}.)

\begin{dialogue}[h]
\begin{quotetable}
\quoteuserline{\p{3}}{RAISED EYEBROWS}
\quotebotline{Raised eyebrows can definitely convey surprise and alertness, which fit perfectly with ``SUDDENLY DANGER.'' ❗ \newline{}\null\newline{} Would you also consider adding a quick, sharp movement with your hands to emulate the suddenness? \newline{} Or perhaps a slight lean forward to convey the immediate attention the situation demands? \newline{}\null\newline{} What do you think would pair well with those raised eyebrows to fully capture the impact? \emojismile{}}
\quoteuserline{\p{3}}{YES}
\quotebotline{Fantastic! Raised eyebrows paired with a quick, sharp hand movement and a slight lean forward should effectively convey ``SUDDENLY DANGER.''\emojidanger{}
}
\end{quotetable}
\caption{Example conversation on \emotingicon{}Emoting.}
\Description{This dialogue illustrates an example conversation on the topic of emoting. Participant P3 suggests using raised eyebrows to convey emotion. The chatbot agrees, noting that raised eyebrows can effectively communicate surprise and alertness, fitting well with the phrase ``SUDDENLY DANGER.'' The bot further suggests adding a quick, sharp hand movement or a slight lean forward to enhance the feeling of immediacy and attention. When P3 responds with agreement, the chatbot confirms that combining raised eyebrows, sharp hand movements, and a slight lean forward should effectively capture the meaning.}
\label{dial:emoting}
\end{dialogue}

Most hearing participants ($N$ = 6) found the \emotingicon{}Emoting conversation valuable for its guidance on facial expressions and body language, helping them to convey emotions effectively. This feature was likened to ``\textit{Surrogacy in ASL} (\p{10})'' and ``\textit{Acting notes from a director} (\p{10})'', with an appreciation for its conceptual approach. 
However, most d/Deaf participants ($N$ = 4), who already internalized sign language's emotional aspects, found the feature less relevant or redundant. 
\p{1} noted, ``\textit{Because I grew up as a Deaf person, I'm already expressive with my face.}''

\bpstart{\timingicon{}Timing}  was used 1.36 times ($SD = 1.12$, $min = 0$, $max = 3$) for \assignedsong{} and 2.1 times ($SD = 2.07$, $min = 0$, $max = 6$) for \selectedsong{} (See \autoref{tab:feature}).
\timingicon{}Timing was rated 2.73 on average ($SD = 1.34$, $min = 1$, $max = 5$), 3.2 by d/Deaf \revised{participants} ($SD = 1.64$), 2.33 by hearing \revised{participants} ($SD = 1.03$) (See \autoref{fig:rating:chat}).
Participants checked the timing of the glosses to align them with the rhythm and tempo of the song (See \autoref{dial:timing}).

\begin{dialogue}[h]
\begin{quotetable}
\quoteuserline{\p{12}}{Yes that sounds great! How about the timing?}
\quotebotline{Great question about timing! Because ASL is so visual, you can match the gloss with the rhythm of the song seamlessly.
\newline{}\null\newline{}
Imagine signing ``MELT'' with a smooth, fluid motion, taking your time to show the melting process, then transitioning into ``SMOOTH'' just as effortlessly. ☕
\newline{}
You can slow down your signs to match the song's tempo. 
\newline{}\null\newline{}
Does that match how you envision it? How do you think your audience will perceive the timing and the flow?} 
\end{quotetable}
\caption{Example conversation on \timingicon{}Timing.}
\Description{This dialogue provides an example of a conversation on the topic of timing. Participant P12 asks about the timing of signs in relation to the song. The chatbot responds by explaining how ASL, being visual, can match the rhythm of the song seamlessly. The bot suggests using smooth, fluid motions for the sign "MELT," taking time to show the process, and then transitioning to "SMOOTH" in a similarly effortless manner, while also adjusting the tempo of the signs to match the song. The chatbot invites P12 to reflect on whether this approach aligns with their vision and how they think the audience will perceive the timing and flow of the signs.}
\label{dial:timing}
\end{dialogue}

Even though both d/Deaf and hearing participants did not discuss \timingicon{}Timing extensively, its explanatory value was appreciated.
Some participants ($N = 5$) found \timingicon{}Timing particularly useful as a 
``\textit{cross-check tool} (\p{7})'', especially when signing along with ``\textit{fast songs} (\p{6})'', to ensure they were on the right track with the rhythm.
The perceived importance of \timingicon{}Timing varied among users. Those with more experience or a natural sense of rhythm often found it less critical, while beginners or those less familiar with musical timing considered it more valuable. As \p{11} noted, ``\textit{[...] it[Timing] is not as crucial as it is for hearing people who are more attuned to musical timing.}''

\subsection{Perceived Attitude and Ownership}

\begin{figure*}
    \centering
\includegraphics[width=\textwidth]{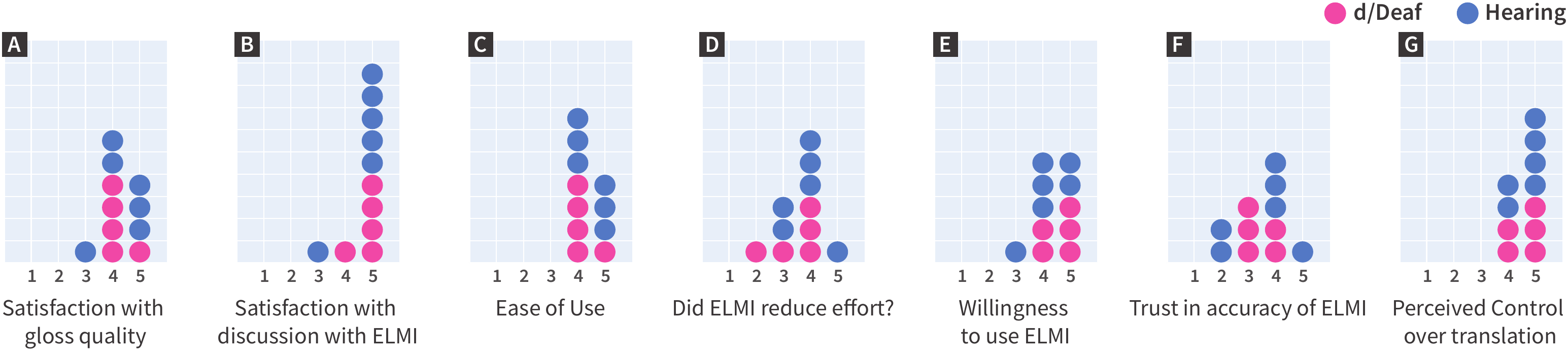}
    \caption{Likert-scale ratings results for quality of gloss, discussion, ease of use, reduction of effort, intention of use, accuracy, and control (1: negative, 5: positive)}
    \Description{This figure presents the post-study Likert-scale ratings results in a histogram format, covering metrics such as the quality of gloss, quality of discussion, ease of use, reduction of effort, intention of use, accuracy, and control. The histogram uses pink circles to represent ratings from Deaf participants and blue circles for hearing participants, showing responses on a scale from 1 (negative) to 5 (positive).}\label{fig:ratings:tam}
\end{figure*}

Reflecting on the conversation with the ELMI chatbot, participants reported various attitudes and stances of the chatbot.
Participants generally found ELMI to be \textbf{encouraging and supportive} in the translation process, helping users feel more confident in their work, while it sometimes provided \textbf{critical and constructive} suggestions helping participants fine-tune their translations. \p{7} and \p{4} noted that ELMI supported balancing between their initial ideas and refined ones: ``\textit{The suggestions helped me align my approach, providing a healthy middle ground if I was overthinking or missing something.} (\p{4})'' 
ELMI was also \textbf{informative and guiding} by providing context that deepened participants' understanding of the songs. \p{7} and \p{10} highlighted how ELMI helped them grasp the essence of the songs, which enhanced their interpretations: ``\textit{There are times when I'm not aware of the context, but ELMI already understands it, guiding me setting the mood, facial expressions, and the story I should convey. Those suggestions are critical.} (\p{7})''

In the post-study survey, participants highly rated their control or authorship over the final gloss, rating 4.63 out of 5 ($SD = 0.5$, $min = 4$, $max$ = 5; see \autoref{fig:ratings:tam}-\blackrectsmall{G}). 
During debriefing, the majority of participants ($N = 11$) echoed this sentiment, noting that they retained control over their work. They appreciated that ELMI did not impose decisions but instead supported their autonomy throughout the translation process.
Participants generally valued their own intuition and experience over ELMI's suggestions, particularly when they disagreed with the tool's input. They used ELMI’s input as a starting point or an additional perspective rather than as a final directive. 
Consistently, participants expressed a strong sense of ownership over their glosses. \p{12} noted, ``\textit{I would definitely give you credit, but it still feels like the work is mostly mine.}''

However, a few participants\delete{ (\p{3}, \p{11})} felt that ELMI sometimes overshadowed their expertise, leading to moments of self-doubt and frustration. \p{3} mentioned feeling discouraged when the AI suggested solutions they had already considered, creating a sense of competition rather than collaboration. Similarly, \p{11} noted that ELMI’s suggestions occasionally did not align with their intended interpretation, causing them to question their own decisions.

\subsection{Strength and Drawbacks of ELMI}

In the post-study survey, participants highly evaluated their satisfaction with the quality of translation they produced using ELMI, rating an average of 4.27 out of 5 on a Likert scale ($SD = 0.65$, $min = 3$\delete{ [\p{9}]}, $max = 5$\delete{ [\p{4}, \p{7}, \p{8}, \p{11}]}).
Participants found ELMI easy to use, giving it an average rating of 4.36 out of 5 ($SD = 0.5$, $min = 4$\delete{ [\p{1}, \p{2}, \p{3}, \p{6}, \p{9}, \p{10}, \p{11}]}, $max = 5$\delete{ [remaining participants]}). d/Deaf participants rated it 4.2 ($SD = 0.45$), while hearing participants rated it slightly higher at 4.5 ($SD = 0.55$) (See \autoref{fig:ratings:tam}-\blackrectsmall{C}). In terms of reducing participants' effort, ELMI received an average rating of 3.63 ($SD = 0.8$, $min = 2$\delete{ [\p{6}]}, $max = 5$\delete{ [\p{9}]}), with d/Deaf participants giving it a 3.4 ($SD = 0.89$) and hearing participants rating it 3.83 ($SD = 0.75$) (See \autoref{fig:ratings:tam}-\blackrectsmall{D}).

However, participants were more critical of ELMI's accuracy, with an average rating of 3.45 out of 5 ($SD = 0.93$, $min = 2$\delete{ [\p{11}, \p{12}]}, $max = 5$\delete{ [\p{9}]}). d/Deaf participants rated the accuracy at 3.4 ($SD = 0.54$), while hearing participants rated it 3.5 ($SD = 1.22$) (See \autoref{fig:ratings:tam}-\blackrectsmall{F}).

\subsubsection{How ELMI Supported the Translation Workflow}

\begin{figure}[b] 
\centering 
\includegraphics[width=\columnwidth]{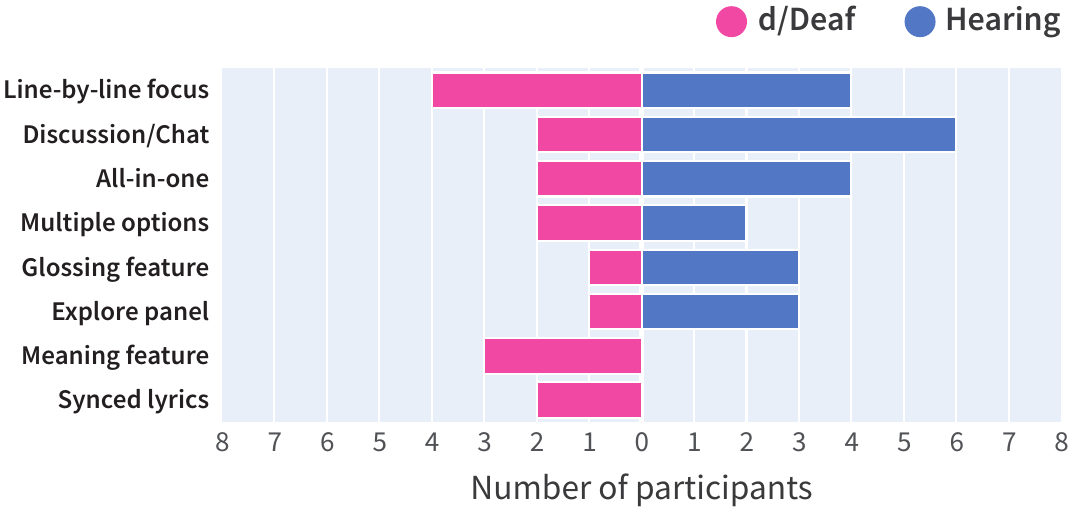} \caption{Distribution of participant responses (39 total) regarding the strengths of ELMI across three choices.}
\Description{This figure shows the distribution of 39 responses regarding the strengths of ELMI. (This is because we asked 13 participants to choose top 3 strengths of ELMI.) The responses are categorized into three different choices, and the distribution is  represented using a bar chart, illustrating how participants rated the key strengths of the system.}

\label{fig:strength} 
\end{figure}

Participants expressed a strong interest in integrating ELMI into their workflow, with an average willingness rating of 4.36 out of 5 on the Likert scale ($SD = 0.67$, $min = 3$\delete{ [\p{11}]}, $max = 5$\delete{ [\p{1}, \p{3}, \p{4}, \p{8}, \p{9}]}). d/Deaf participants rated their willingness slightly higher, at 4.6 ($SD = 0.55$), while hearing participants rated it 4.17 ($SD = 0.75$).
\autoref{fig:strength} summarizes the main features of ELMI that participants mentioned as the top three favorites in the post-study surveys. In the following, we cover the noteworthy features.

\ipstart{Line-by-line Focus}
Most of the participants ($N$ = 8, 4 d/Deaf and 4 hearing) liked ELMI's ability to break down song lyrics into ``lines'', making the translation process more precise (See \autoref{fig:strength}). This was particularly useful for complex lyrics, where understanding the meaning is crucial. 
Additionally, looping specific lines with the video was received as useful for practice and refinement. \p{11} mentioned ELMI's ``\textit{line-by-line setup made it easy to see how the person in the video is conveying emotions and actions, which is incredibly helpful for aligning translation with the intended expression.}''

\ipstart{Interactive Discussion and Chat}
As \p{9} described, ``\textit{It doesn’t feel like you’re alone in interpreting}'', discussing with ELMI was highly rated by participants ($N$ = 8, 2 d/Deaf and 6 hearing) for providing a sense of independence and efficiency during the translation process, reducing their reliance on other people for feedback (See \autoref{fig:strength}). \p{2} remarked, ``\textit{[...] Without ELMI, I’d have to ask my interpreter how the voice sounds, which can be a long process. It really depends on the person I’m working with, and I have to trust them more. But ELMI gives me more confidence and independence.}''
Although \p{13} also noted that it may not entirely replace the need for direct consultation when facing difficult decisions, \p{10} mentioned that ELMI serves as an effective ``intermediary tool,'' helping to refine initial thoughts before seeking further input from others making the process more thorough and reflective.

\ipstart{ALL-IN-ONE workplace}
The integration of lyrics, glosses, videos, and contextual information into a single platform was widely seen as a major advantage ($N$ = 6, 2 d/Deaf and 4 hearing), streamlining the workflow and allowing participants to focus more on the translation itself (See \autoref{fig:strength}).
\p{4} highlighted the convenience of this approach: ``\textit{Normally, I would have to navigate between different resources, which can be cumbersome. If I needed to verify something, I had to keep going back and forth.}''

\ipstart{Synced Lyrics with Music Video}
Deaf participants\delete{ (\p{1}, \p{2}, \p{3})} particularly valued ELMI's ability to synchronize lyrics with the music, making the process of matching the song’s tempo more accessible.
Usually, this process involved filming themselves, using metronomes, and consulting with interpreters to ensure their sync was accurate. However, ELMI simplified this for deaf participants.
\p{2} highlighted the effectiveness of the visual cues: 
``\textit{I like the highlighted words in red that pop up. With the cues, I can easily follow the beat and see which word is currently being emphasized. As a deaf person, it’s very helpful to see those and understand the song’s tempo.}''
\p{1} referred to this feature as ``Karaoke methods'': ``\textit{It allows me to see where the song is sung. I'm profoundly deaf and wanted to know this!}''

\subsubsection{Challenges and Concerns of ELMI}
Besides the benefits, participants also encountered challenges and shared various concerns about ELMI regarding the d/Deaf community and culture.

\ipstart{Loss of Human Interaction and Cultural Appropriation}
Some participants (\p{3}, \p{7}) expressed concern that ELMI diminishes the ``personal touch of human interaction (\p{3})''. They missed the immediate feedback and personalized discussions that typically come from working with other people. Additionally, d/Deaf participants\delete{ (\p{2}, \p{3}, \p{5})} voiced concern that hearing individuals using ELMI might unintentionally engage in cultural appropriation, urging for careful usage: ``\textit{I think interpreting songs really needs to spotlight the Deaf community. If a hearing person uses technology like ELMI, it might lead to cultural appropriation. They might use it for their own benefit, without the intention of spreading sign language or supporting the Deaf community.}'' (\p{2})

Participants suggested ways to ensure accuracy and help maintain the human element in interpreting songs.
d/Deaf participants (\p{2}, \p{3}) suggested collaboration features to share their work with interpreters, take notes, and get feedback from others to refine their signs.
Additionally, \p{10} emphasized the importance of adding a disclaimer: \textit{``This tool doesn't replace the expertise of a Deaf person. When signing music, it's important to consult with a Deaf person, even when using this tool.''}, reminding users to consult with Deaf individuals during the translation process. 
Furthermore, \p{11} highlighted the need to integrate feedback from the Deaf community regarding the use of AI in conveying nuanced expressions.
``\textit{The effectiveness of AI in conveying expressions is still uncertain. I believe a real person could convey these nuances better than AI. It might be useful to survey the Deaf community to understand their preferences whether they'd prefer real human demonstrations or if they'd be comfortable with AI-generated visuals.}''

\ipstart{English-centric Generation hampering Accuracy and Depth of Translation}
ELMI was seen to be ``English-focused (\p{3})'', with its concepts presented from a ``non-Deaf perspective (\p{10})''. 
Participants (\p{3}, \p{10}, \p{13}) noted that ELMI might be better suited for Pidgin Sign Language (PSE) or Signed Exact English (SEE).
\p{13} noticed ELMI often adhered English word order, lacking deeper integration with ASL grammar and nuances: ``\textit{I’m noticing that ELMI's suggestions for the gloss are still following English word order, even though it was set to ASL. I understand that it's challenging for AI to grasp ASL grammar. For example, using signs like 'stealing person' and the concept of 'undercover' seems more like PSE than ASL.}''

Additionally, ELMI's responses were sometimes seen as too shallow, particularly when dealing with poetic expressions. Participants (\p{9}, \p{11}, \p{13}) felt that ELMI often failed to fully grasp the deeper meanings behind certain phrases. \p{11} concern that ELMI might not fully capture the nuances, ``\textit{I'm unsure if ELMI can handle such specific phrases without seeing them in action.}'', given that ELMI is not actually observing their performance.
\p{12} also questioned the accuracy of ASL suggestions that ELMI might not be drawing from sign dictionaries: ``\textit{It may be suggesting valid glossing, but also could be hallucinating.}'' This uncertainty led to a lack of trust in ELMI's recommendations.

To make ELMI more culturally and linguistically appropriate, participants offered several key suggestions. Many participants ($N = 7$) recommended integrating an ASL dictionary that includes image or video demonstrations, as well as detailed hand shapes, which would be especially beneficial for novice song-signers. For instance, \p{3} remarked, ``\textit{Seeing examples of signs could bridge the gap for students who are still learning ASL. For example, if I could click on something and see the sign for a word like `butter' or `smooth,' it would be very helpful.}'' 


\ipstart{Lose of Context due to Line-by-line Focus}
Although the concept of line-by-line focused translation was generally well-received, some participants\delete{ (\p{5}, \p{9}, \p{10}, \p{12}, \p{13})} raised concerns on it; they felt focusing on a line basis could overlook the broader context of the song and disrupt the continuity of ideas. As \p{5} mentioned: ``\textit{If someone were using ELMI, they might focus too much on translating line by line, instead of considering the overall structure.}'' 
As such, \p{4} and \p{5} suggested supporting multiple-line selection, which allows users to consider the broader context of the song, addressing the narrow focus of the line-by-line approach. \p{4} explained that for slower songs, the current single-line view works well. But for faster genres like hip hop, viewing two lines at once would help the user better understand connections between lines, such as punchlines and their surrounding context. 

\ipstart{Insufficient Critical Feedback} 
Four participants\delete{ (\p{3}, \p{5}, \p{11}, \p{12})} pointed out that ELMI was sometimes seen as too supportive, often providing redundant feedback that did not challenge the participants' interpretations. 
This led to concerns about the tool not being critical enough, which could hinder improvement. Participants (\p{3}, \p{5}) also worried that less experienced signers might misuse ELMI’s suggestions, potentially adopting incorrect interpretations. As \p{3} noted, ``\textit{This could serve as a shortcut, even though [they] haven't fully studied.}''
\section{Discussion}

In this section, we discuss lessons learned from the design and evaluation of ELMI. We also reflect on the implications for better supporting artistic sign language translation in an accessible and culturally sensitive manner.

\subsection{Tailoring Feedback for d/Deaf and Hearing Song-Signers}
\delete{Our study uncovered participants' preferences for discussion topics, which participants rated as useful and frequently used in the following order: \glossingicon{}glossing, \meaningicon{}meaning, \emotingicon{}emoting, and \timingicon{}timing. 
However, there were} \revised{We found} notable differences in how d/Deaf and hearing song-signers prioritized \revised{the four discussion topics}.
d/Deaf song-signers placed stronger importance on \timingicon{}timing, aligning their signs with the beat and rhythm of the music. On the other hand, hearing song-signers placed greater importance on \emotingicon{}emoting, conveying the emotional content of the lyrics. This aligns with Maler’s findings that d/Deaf and hearing song-signers have distinct approaches to using physical space and rhythmic techniques \cite{maler2015musical}: Hearing signers are often more concerned with replicating the pitches and rhythms of the music through body movement, while d/Deaf signers pay closer attention to the relationship between the rhythm of the music and the prosody of ASL. 

Yoo \etal{} highlight this dynamic that d/Deaf and hearing song-signers have ``complementary'' strengths and weaknesses \cite{yoo2023understanding}. Our study aligned with these findings, particularly in expressive and rhythmic translations.
Hearing participants reported struggling to convey the expressive nature of ASL, while d/Deaf participants felt more naturally expressive. 
In contrast, d/Deaf song-signers (especially those profoundly deaf) shared that timing was a significant challenge, requiring substantial effort to align their signs with the music’s beat. 
Hearing participants, who were more attuned to the timing and rhythm of music, found this less of a concern.
Some d/Deaf participants (\p{2}, \p{5}) expressed interest in additional timing aids, such as a metronome, to enhance their synchronization with the music.
Thus, it is crucial to tailor feedback to address the distinct needs of each group, allowing both d/Deaf and hearing song-signers to build on their strengths while overcoming their specific challenges.


\subsection{Enhancing Cultural Sensitivity and Awareness of the Tool}
The integration of AI tools like ELMI in ASL translation brings potential benefits but also requires careful attention to cultural sensitivity and ethical considerations.
ELMI received positive feedback for its discussion feature, which led to more confident and independent translations. 
However, some participants worried that users might become overly reliant on ELMI, potentially overlooking the value of human feedback. 
Additionally, d/Deaf participants raised concerns that ELMI could unintentionally engage in cultural appropriation or misuse in ways that may harm the Deaf community. 
\revised{These concerns are also applicable to other LLM-infused features, such as the emotional and performance guide on the Information Panel. While these features were intended to aid users in creating nuanced and visually expressive ASL translations, participants highlighted the need to ensure that such AI-generated suggestions are culturally appropriate and do not inadvertently misrepresent Deaf culture or ASL conventions. As \p{10} noted, consulting with Deaf individuals during the development of such tools is crucial for ensuring cultural alignment and personalization of the system.}

Therefore, future work would be needed to understand the perspectives of Deaf individuals on using AI for ASL translations. Following \p{11}'s suggestion, \textit{``It might be useful to survey the Deaf community to understand their preferences whether they'd prefer real human demonstrations or if they'd be comfortable with AI-generated visuals.''}
Furthermore, \p{12} pointed out that some of ELMI’s responses seemed unreliable, suggesting potential hallucinations, as it was unclear whether the LLM had been trained with accurate ASL signs. 
To address this, we suggest augmenting the model training dataset with ASL gloss \cite{moryossef2021data, cao2022explore} and text pairs produced by certified interpreters. Additionally, we propose that an extra layer of inference can be incorporated to validate the quality of cultural nuances and inspect any implicit biases or mischaracterization of the Deaf context ingrained in the gloss, ensuring more reliable translations.

\subsection{Expanding Scope beyond Single Lines}
The line-by-line focus unit was one of the strengths of the ELMI that both d/Deaf and hearing song-signers appreciated. However, some participants expressed concerns that focusing on individual lines might be too narrow and could disrupt the overall cohesiveness and flow of the song. The unit is currently fixed to the line reference from Genius lyrics, but since the unit users want to discuss with ELMI at one time may differ, it needs to be customizable.
Participants suggested that the appropriate focus should depend on the complexity of the song. For example, hip-hop often requires analyzing multiple lines together to understand the connection between punchlines (\p{4}). 
Verses typically advance the narrative with new lyrics, while choruses tend to feature memorable, repeated content \cite{summach2011structure}. Additionally, features extracted from verses and choruses can have different predictive powers in genre recognition tasks \cite{mayerl2022verse}.
Therefore, future song-signing tools should support discussion with an arbitrary range of the context. One option is to allow users to select blocks of text or lines, depending on the song’s genre and structure. A drag-and-drop feature, similar to the commenting system in Google Docs \cite{morse2021increase, roberts2013chalk}, would give users the ability to discuss larger units of content.

\subsection{Shaping Tone and Persona of an AI}
The stance of AI in artistic ASL translation requires careful consideration. ELMI was described as ``encouraging and supportive” that enhanced participants' translations, while others saw it as “informative and guiding” that facilitated their work. Although some song-signers found ELMI’s feedback to be ``critical and constructive,” helping them identify areas for improvement in their glossing, others felt it was ``too agreeable.” 
They expressed concern that this might lead to misuse or over-reliance on ELMI, as the tool's feedback may not be critical enough for meaningful improvement.
This feedback may have been influenced by the prompt, which stated ELMI to be supportive and encourage critical thinking (See Appendix). 
Future work should focus on prompt engineering to strike a balance between supportive and critical feedback, as this balance impacts the trustworthiness of the results \cite{kim2024m}. 
Research shows that the tone of AI advice influences user perceptions. Advice given in a `skeptical’ style is perceived as more unpredictable, while advice in a `whimsical’ style is seen as less malicious \cite{wester2024exploring}. 
Therefore, it is essential to refine ELMI’s characteristics to achieve the right balance between encouragement and constructive critique to optimize user trust and performance.

\subsection{Supporting Rich Recording Modalities beyond Glossing}
Our gloss analysis indicates that d/Deaf individuals use more non-manual signals (NMS) than hearing song-signers, including classifiers and visual vernacular. This confirms Maler’s findings that d/Deaf song-signers embody music differently from hearing signers by creating a visual, kinetic form of music in sign language, rather than using sign language to express acoustic elements \cite{maler2015musical}. 
Additionally, while d/Deaf individuals tend to use ASL with NMS, hearing individuals are more likely to use Pidgin Signed English (PSE) or Signed Exact English (SEE). Hearing signers often rely on SEE because they are less familiar with the grammatical structure of ASL and may struggle to accurately translate English into ASL~\cite{maler2015musical}. 
In this sense, our system's emphasis on \emotingicon{}emoting and \timingicon{}timing may not fully benefit novice song-signers who lack a strong foundation in ASL grammar. 
Therefore, it is important to consider the user’s proficiency level and fine-tune feedback accordingly.

We also found that many d/Deaf individuals do not use glossing in their own contexts. Although most participants were familiar with the concept of glossing, d/Deaf song-signers preferred to practice using video recordings rather than glossing. 
Captions using glosses were more frequently observed among hearing song-signers \cite{maler2015musical}. 
One limitation of glossing is that it lacks a standardized form and is more commonly used by hearing individuals to learn ASL \cite{farooq2021advances}. 
To better support full ASL input, systems can incorporate video recordings and editing. Future work could involve using video input for ASL recognition \cite{liu2024improving, pigou2015sign}, translating the signs into gloss, and then inputting them into a large language model (LLM). This would enable ELMI to observe and analyze the signs and expressions to give better feedback.

\subsection{\revised{Limitations and Future Work}}
Even though we made efforts to balance the participation of d/Deaf and hearing individuals to capture perspectives from both communities, the majority of our participants ($N = 7$) are hearing. 
Although similar studies often focused on hearing and hard of hearing song-signers \cite{yoo2023understanding} and hearing song-signers are reported to be abundant on social media platforms \cite{maler2015musical}, our participant pool may not fully represent the entire online song-signer population. \revised{This work is just the first step toward designing and developing an accessible song-signing tool, and we learned that song-signers could successfully adopt the concept of ELMI. \cameraready{To further enhance the design of nuanced features, such as video-based support, future AI-assisted tools could integrate direct ASL dictionary access with example signing videos and curated performance references. Additionally,} future work may consider adopting participatory design approaches~\cite{quintero2022review, schuler1993participatory} that directly involve song signers throughout the development process.}

Our study focused on analyzing \assignedsong{}, BTS's ``Butter,''  a Dance-pop track.
While pop music is one of the most common and popular genres \cite{araujo2018identification, holt2019genre}, we acknowledge that the song may not be aligned with some participants' personal interests or expertise regarding its genre and style. Nonetheless, \assignedsong{} allowed consistency in comparison across participants. 
To complement this, we also involved \selectedsong{}, providing participants with a more personalized experience. For example, some participants (\p{1}, \p{6}, \p{11}), who were involved in church worship, selected religious songs as their \selectedsong{}.

We also note that ELMI's performance in lyric analysis and discussion may vary across different popularity of songs. Since ELMI relies on databases like Genius and YouTube’s closed captioning for lyric processing, the model may struggle with songs that are less popular or lack robust reference data. 
Well-documented songs are likely to yield richer feedback, but custom or niche tracks may result in less detailed explanations or feedback. \revised{Thus, ELMI's ability to address cultural, regional, or genre-specific nuances remains uncertain.} Future research could explore the limitations of LLMs in processing less-known songs and assess their adaptability across varying musical genres.

\revised{Lastly, we acknowledge that GPT-4.0 \cite{openai2023gpt4} without fine-tuning for sign-language-specific tasks, such as generating ASL glosses, might present concerns about its accuracy and applicability. 
While the formative study and prior work \cite{shahin2023chatgpt, othman2024comparative} highlighted GPT's potential to support sign language translation, its ability to produce accurate and linguistically robust glosses remains unproven.
Future work can incorporate expert evaluations with song-signers to assess the quality and applicability of GPT-generated glosses.} 
\section{Conclusion}
In this paper, we presented ELMI, an accessible, LLM-driven system designed to assist song-signers in translating lyrics into sign language. 
Informed by our formative study, ELMI was developed to support semantic, syntactic, expressive, and rhythmic aspects of translation. 
We found that ELMI enriched the translation process by providing visual cues, interactive discussions, and individual line focus. 
All participants successfully integrated ELMI into their workflow, gaining increased confidence and independence, with control and ownership in their glossing process.
We recommend tailoring feedback to meet the needs of both d/Deaf and hearing song-signers, enhancing the system's cultural sensitivity, broadening its focus to multiline translations, refining prompts for more critical discussions, and supporting diverse inputs.
We hope ELMI will inspire and inform researchers in creativity support and accessibility, encouraging the use of LLMs to foster culturally sensitive and accessible tools.

\begin{acks}
We thank our study participants in both the formative and main studies for their time and efforts. We are also grateful to Dasom Choi for her feedback on the early version of our draft. Inhwa Song and Ken Christofferson gave us feedback on our prototype. This work was supported through a research internship at NAVER AI Lab of NAVER Cloud and in part by NSERC grant RGPIN-2021-04268.
\end{acks}

\bibliographystyle{ACM-Reference-Format}
\bibliography{bibliography}


\begin{thebibliography}{76}


\ifx \showCODEN    \undefined \def \showCODEN     #1{\unskip}     \fi
\ifx \showDOI      \undefined \def \showDOI       #1{#1}\fi
\ifx \showISBNx    \undefined \def \showISBNx     #1{\unskip}     \fi
\ifx \showISBNxiii \undefined \def \showISBNxiii  #1{\unskip}     \fi
\ifx \showISSN     \undefined \def \showISSN      #1{\unskip}     \fi
\ifx \showLCCN     \undefined \def \showLCCN      #1{\unskip}     \fi
\ifx \shownote     \undefined \def \shownote      #1{#1}          \fi
\ifx \showarticletitle \undefined \def \showarticletitle #1{#1}   \fi
\ifx \showURL      \undefined \def \showURL       {\relax}        \fi
\providecommand\bibfield[2]{#2}
\providecommand\bibinfo[2]{#2}
\providecommand\natexlab[1]{#1}
\providecommand\showeprint[2][]{arXiv:#2}

\bibitem[Amin et~al\mbox{.}(2021)]%
        {amin2021sign}
\bibfield{author}{\bibinfo{person}{Mohamed Amin}, \bibinfo{person}{Hesahm Hefny}, {and} \bibinfo{person}{Mohammed Ammar}.} \bibinfo{year}{2021}\natexlab{}.
\newblock \showarticletitle{Sign language gloss translation using deep learning models}.
\newblock \bibinfo{journal}{\emph{International Journal of Advanced Computer Science and Applications}} \bibinfo{volume}{12}, \bibinfo{number}{11} (\bibinfo{year}{2021}).
\newblock


\bibitem[Andayani et~al\mbox{.}(2023)]%
        {andayani2023inquiry}
\bibfield{author}{\bibinfo{person}{Andayani Andayani}, \bibinfo{person}{Herman Herman}, \bibinfo{person}{Isry~Laila Syathroh}, \bibinfo{person}{Endang Fatmawati}, \bibinfo{person}{Ninawati Syahrul}, \bibinfo{person}{Nabil Al-Awawdeh}, \bibinfo{person}{Junita Batubara}, {and} \bibinfo{person}{Nanda Saputra}.} \bibinfo{year}{2023}\natexlab{}.
\newblock \showarticletitle{Inquiry into the challenges of translating idioms extracted from musical lyrics}.
\newblock \bibinfo{journal}{\emph{Research Journal in Advanced Humanities}} \bibinfo{volume}{4}, \bibinfo{number}{3} (\bibinfo{year}{2023}).
\newblock


\bibitem[Apter and Herman(2016)]%
        {apter2016translating}
\bibfield{author}{\bibinfo{person}{Ronnie Apter} {and} \bibinfo{person}{Mark Herman}.} \bibinfo{year}{2016}\natexlab{}.
\newblock \bibinfo{booktitle}{\emph{Translating for singing: The theory, art and craft of translating lyrics}}.
\newblock \bibinfo{publisher}{Bloomsbury publishing}.
\newblock


\bibitem[Araujo and Nakamura(2018)]%
        {araujo2018identification}
\bibfield{author}{\bibinfo{person}{Carlos~VS Araujo} {and} \bibinfo{person}{Eduardo~F Nakamura}.} \bibinfo{year}{2018}\natexlab{}.
\newblock \showarticletitle{Identification of most popular musical genres and their influence factors}. In \bibinfo{booktitle}{\emph{Proceedings of the 24th Brazilian Symposium on Multimedia and the Web}}. \bibinfo{pages}{233--236}.
\newblock


\bibitem[Aryawan et~al\mbox{.}(2022)]%
        {aryawan2022deixis}
\bibfield{author}{\bibinfo{person}{LPKF Aryawan}, \bibinfo{person}{IG Budasi}, {and} \bibinfo{person}{DP Ramendra}.} \bibinfo{year}{2022}\natexlab{}.
\newblock \showarticletitle{The deixis used in the song lyrics performed by the most popular boy group BTS}.
\newblock \bibinfo{journal}{\emph{Jurnal Pendidikan Bahasa Inggris Indonesia}} \bibinfo{volume}{10}, \bibinfo{number}{1} (\bibinfo{year}{2022}), \bibinfo{pages}{30--39}.
\newblock


\bibitem[Athitsos et~al\mbox{.}(2008)]%
        {Vassilis2008ASLLVD}
\bibfield{author}{\bibinfo{person}{Vassilis Athitsos}, \bibinfo{person}{Carol Neidle}, \bibinfo{person}{Stan Sclaroff}, \bibinfo{person}{Joan Nash}, \bibinfo{person}{Alexandra Stefan}, \bibinfo{person}{Quan Yuan}, {and} \bibinfo{person}{Ashwin Thangali}.} \bibinfo{year}{2008}\natexlab{}.
\newblock \showarticletitle{The American Sign Language Lexicon Video Dataset}. In \bibinfo{booktitle}{\emph{2008 IEEE Computer Society Conference on Computer Vision and Pattern Recognition Workshops}}. \bibinfo{pages}{1--8}.
\newblock
\urldef\tempurl%
\url{https://doi.org/10.1109/CVPRW.2008.4563181}
\showDOI{\tempurl}


\bibitem[Bahan(2006)]%
        {bahan2006face}
\bibfield{author}{\bibinfo{person}{Ben Bahan}.} \bibinfo{year}{2006}\natexlab{}.
\newblock \showarticletitle{Face-to-face tradition in the American Deaf Community}.
\newblock \bibinfo{journal}{\emph{Signing the body poetic}} (\bibinfo{year}{2006}), \bibinfo{pages}{21--50}.
\newblock


\bibitem[Blom et~al\mbox{.}(2014)]%
        {blom2014finding}
\bibfield{author}{\bibinfo{person}{Helen Blom}, \bibinfo{person}{Marc Marschark}, \bibinfo{person}{Mathijs~PJ Vervloed}, {and} \bibinfo{person}{Harry Knoors}.} \bibinfo{year}{2014}\natexlab{}.
\newblock \showarticletitle{Finding friends online: Online activities by deaf students and their well-being}.
\newblock \bibinfo{journal}{\emph{PloS one}} \bibinfo{volume}{9}, \bibinfo{number}{2} (\bibinfo{year}{2014}), \bibinfo{pages}{e88351}.
\newblock


\bibitem[Braun and Clarke(2012)]%
        {braun2012thematic}
\bibfield{author}{\bibinfo{person}{Virginia Braun} {and} \bibinfo{person}{Victoria Clarke}.} \bibinfo{year}{2012}\natexlab{}.
\newblock \bibinfo{booktitle}{\emph{Thematic analysis.}}
\newblock \bibinfo{publisher}{American Psychological Association}.
\newblock


\bibitem[Braun and Clarke(2019)]%
        {braun2019reflecting}
\bibfield{author}{\bibinfo{person}{Virginia Braun} {and} \bibinfo{person}{Victoria Clarke}.} \bibinfo{year}{2019}\natexlab{}.
\newblock \showarticletitle{Reflecting on reflexive thematic analysis}.
\newblock \bibinfo{journal}{\emph{Qualitative research in sport, exercise and health}} \bibinfo{volume}{11}, \bibinfo{number}{4} (\bibinfo{year}{2019}), \bibinfo{pages}{589--597}.
\newblock


\bibitem[Br{\'e}t{\'e}ch{\'e} and Esclapez(2018)]%
        {breteche2018music}
\bibfield{author}{\bibinfo{person}{Sylvain Br{\'e}t{\'e}ch{\'e}} {and} \bibinfo{person}{Christine Esclapez}.} \bibinfo{year}{2018}\natexlab{}.
\newblock \showarticletitle{Music (s), Musicology and Science: Towards an Interscience Network: The Example of the Deaf Musical Experience}. In \bibinfo{booktitle}{\emph{Music Technology with Swing: 13th International Symposium, CMMR 2017, Matosinhos, Portugal, September 25-28, 2017, Revised Selected Papers 13}}. Springer, \bibinfo{pages}{637--657}.
\newblock


\bibitem[Camgoz et~al\mbox{.}(2018)]%
        {camgoz2018neural}
\bibfield{author}{\bibinfo{person}{Necati~Cihan Camgoz}, \bibinfo{person}{Simon Hadfield}, \bibinfo{person}{Oscar Koller}, \bibinfo{person}{Hermann Ney}, {and} \bibinfo{person}{Richard Bowden}.} \bibinfo{year}{2018}\natexlab{}.
\newblock \showarticletitle{Neural Sign Language Translation}. In \bibinfo{booktitle}{\emph{IEEE Conference on Computer Vision and Pattern Recognition (CVPR)}}.
\newblock


\bibitem[Camgoz et~al\mbox{.}(2020)]%
        {camgoz2020sign}
\bibfield{author}{\bibinfo{person}{Necati~Cihan Camgoz}, \bibinfo{person}{Oscar Koller}, \bibinfo{person}{Simon Hadfield}, {and} \bibinfo{person}{Richard Bowden}.} \bibinfo{year}{2020}\natexlab{}.
\newblock \showarticletitle{Sign language transformers: Joint end-to-end sign language recognition and translation}. In \bibinfo{booktitle}{\emph{Proceedings of the IEEE/CVF conference on computer vision and pattern recognition}}. \bibinfo{pages}{10023--10033}.
\newblock


\bibitem[Cao et~al\mbox{.}(2022)]%
        {cao2022explore}
\bibfield{author}{\bibinfo{person}{Yong Cao}, \bibinfo{person}{Wei Li}, \bibinfo{person}{Xianzhi Li}, \bibinfo{person}{Min Chen}, \bibinfo{person}{Guangyong Chen}, \bibinfo{person}{Long Hu}, \bibinfo{person}{Zhengdao Li}, {and} \bibinfo{person}{Kai Hwang}.} \bibinfo{year}{2022}\natexlab{}.
\newblock \showarticletitle{Explore More Guidance: A Task-aware Instruction Network for Sign Language Translation Enhanced with Data Augmentation}. In \bibinfo{booktitle}{\emph{Findings of the Association for Computational Linguistics: NAACL 2022}}.
\newblock


\bibitem[Choi(2018)]%
        {choi2018computational}
\bibfield{author}{\bibinfo{person}{Kahyun Choi}.} \bibinfo{year}{2018}\natexlab{}.
\newblock \emph{\bibinfo{title}{Computational lyricology: quantitative approaches to understanding song lyrics and their interpretations}}.
\newblock \bibinfo{thesistype}{Ph.\,D. Dissertation}. \bibinfo{school}{University of Illinois at Urbana-Champaign}.
\newblock


\bibitem[Cripps et~al\mbox{.}(2019)]%
        {cripps2019signed}
\bibfield{author}{\bibinfo{person}{Jody Cripps}, \bibinfo{person}{Ely Rosenblum}, {and} \bibinfo{person}{Anita Small}.} \bibinfo{year}{2019}\natexlab{}.
\newblock \showarticletitle{Signed music: An emerging interperformative art}.
\newblock \bibinfo{journal}{\emph{Waypoint: Proceedings of the Sixth Biennial Deaf Studies Today}} (\bibinfo{year}{2019}).
\newblock


\bibitem[Cripps(2017)]%
        {crippsunderstanding}
\bibfield{author}{\bibinfo{person}{Jody~H Cripps}.} \bibinfo{year}{2017}\natexlab{}.
\newblock \showarticletitle{Understanding Signed Music Jody H. Cripps Ely Lyonblum}.
\newblock \bibinfo{journal}{\emph{Society for American Sign Language Journal}} \bibinfo{volume}{1}, \bibinfo{number}{1} (\bibinfo{year}{2017}).
\newblock


\bibitem[Davis et~al\mbox{.}(1989)]%
        {davis1989technology}
\bibfield{author}{\bibinfo{person}{Fred~D Davis}, \bibinfo{person}{RP Bagozzi}, {and} \bibinfo{person}{PR Warshaw}.} \bibinfo{year}{1989}\natexlab{}.
\newblock \showarticletitle{Technology acceptance model}.
\newblock \bibinfo{journal}{\emph{J Manag Sci}} \bibinfo{volume}{35}, \bibinfo{number}{8} (\bibinfo{year}{1989}), \bibinfo{pages}{982--1003}.
\newblock


\bibitem[Desai et~al\mbox{.}(2024)]%
        {desai2024systemic}
\bibfield{author}{\bibinfo{person}{Aashaka Desai}, \bibinfo{person}{Maartje De~Meulder}, \bibinfo{person}{Julie~A Hochgesang}, \bibinfo{person}{Annemarie Kocab}, {and} \bibinfo{person}{Alex~X Lu}.} \bibinfo{year}{2024}\natexlab{}.
\newblock \showarticletitle{Systemic Biases in Sign Language AI Research: A Deaf-Led Call to Reevaluate Research Agendas}.
\newblock \bibinfo{journal}{\emph{Proceedings of the 11th Workshop on the Representation and Processing of Sign Languages, pages 54–65}} (\bibinfo{year}{2024}).
\newblock


\bibitem[Emmorey et~al\mbox{.}(1993)]%
        {emmorey1993visual}
\bibfield{author}{\bibinfo{person}{Karen Emmorey}, \bibinfo{person}{Stephen~M Kosslyn}, {and} \bibinfo{person}{Ursula Bellugi}.} \bibinfo{year}{1993}\natexlab{}.
\newblock \showarticletitle{Visual imagery and visual-spatial language: Enhanced imagery abilities in deaf and hearing ASL signers}.
\newblock \bibinfo{journal}{\emph{Cognition}} \bibinfo{volume}{46}, \bibinfo{number}{2} (\bibinfo{year}{1993}), \bibinfo{pages}{139--181}.
\newblock


\bibitem[Farooq et~al\mbox{.}(2021)]%
        {farooq2021advances}
\bibfield{author}{\bibinfo{person}{Uzma Farooq}, \bibinfo{person}{Mohd Shafry~Mohd Rahim}, \bibinfo{person}{Nabeel Sabir}, \bibinfo{person}{Amir Hussain}, {and} \bibinfo{person}{Adnan Abid}.} \bibinfo{year}{2021}\natexlab{}.
\newblock \showarticletitle{Advances in machine translation for sign language: approaches, limitations, and challenges}.
\newblock \bibinfo{journal}{\emph{Neural Computing and Applications}} \bibinfo{volume}{33}, \bibinfo{number}{21} (\bibinfo{year}{2021}), \bibinfo{pages}{14357--14399}.
\newblock


\bibitem[Fauqi and Eva(2020)]%
        {fauqi2020metaphorical}
\bibfield{author}{\bibinfo{person}{Izzet~Fikrotul Fauqi} {and} \bibinfo{person}{Eva Eva}.} \bibinfo{year}{2020}\natexlab{}.
\newblock \showarticletitle{Metaphorical Entailment in BTS Song Lyrics: A Semantic Perspective}.
\newblock \bibinfo{journal}{\emph{Proceedings of The ICECRS}}  \bibinfo{volume}{8} (\bibinfo{year}{2020}).
\newblock


\bibitem[Fisher(2021)]%
        {fisher2021embodied}
\bibfield{author}{\bibinfo{person}{Vicky~J Fisher}.} \bibinfo{year}{2021}\natexlab{}.
\newblock \showarticletitle{Embodied songs: Insights into the nature of cross-modal meaning-making within sign language informed, embodied interpretations of vocal music}.
\newblock \bibinfo{journal}{\emph{Frontiers in psychology}}  \bibinfo{volume}{12} (\bibinfo{year}{2021}), \bibinfo{pages}{624689}.
\newblock


\bibitem[Franzon(2008)]%
        {franzon2008choices}
\bibfield{author}{\bibinfo{person}{Johan Franzon}.} \bibinfo{year}{2008}\natexlab{}.
\newblock \showarticletitle{Choices in song translation: Singability in print, subtitles and sung performance}.
\newblock \bibinfo{journal}{\emph{The Translator}} \bibinfo{volume}{14}, \bibinfo{number}{2} (\bibinfo{year}{2008}), \bibinfo{pages}{373--399}.
\newblock


\bibitem[Gong et~al\mbox{.}(2024)]%
        {gong2024llms}
\bibfield{author}{\bibinfo{person}{Jia Gong}, \bibinfo{person}{Lin~Geng Foo}, \bibinfo{person}{Yixuan He}, \bibinfo{person}{Hossein Rahmani}, {and} \bibinfo{person}{Jun Liu}.} \bibinfo{year}{2024}\natexlab{}.
\newblock \showarticletitle{Llms are good sign language translators}. In \bibinfo{booktitle}{\emph{Proceedings of the IEEE/CVF Conference on Computer Vision and Pattern Recognition}}. \bibinfo{pages}{18362--18372}.
\newblock


\bibitem[Gradistina(2023)]%
        {gradistina2023figurative}
\bibfield{author}{\bibinfo{person}{Luh Komang~Mira Gradistina}.} \bibinfo{year}{2023}\natexlab{}.
\newblock \emph{\bibinfo{title}{Figurative Language Used in BTS Selected Song Lyrics}}.
\newblock \bibinfo{thesistype}{Ph.\,D. Dissertation}. \bibinfo{school}{Universitas Mahasaraswati Denpasar}.
\newblock


\bibitem[Guardino and Cannon(2016)]%
        {guardino2016deafness}
\bibfield{author}{\bibinfo{person}{Caroline Guardino} {and} \bibinfo{person}{Joanna~E Cannon}.} \bibinfo{year}{2016}\natexlab{}.
\newblock \showarticletitle{Deafness and diversity: Reflections and directions}.
\newblock \bibinfo{journal}{\emph{American Annals of the Deaf}} \bibinfo{volume}{161}, \bibinfo{number}{1} (\bibinfo{year}{2016}), \bibinfo{pages}{104--112}.
\newblock


\bibitem[Holt(2019)]%
        {holt2019genre}
\bibfield{author}{\bibinfo{person}{Fabian Holt}.} \bibinfo{year}{2019}\natexlab{}.
\newblock \bibinfo{booktitle}{\emph{Genre in popular music}}.
\newblock \bibinfo{publisher}{University of Chicago Press}.
\newblock


\bibitem[Hubscher-Davidson(2017)]%
        {hubscher2017translation}
\bibfield{author}{\bibinfo{person}{S{\'e}verine Hubscher-Davidson}.} \bibinfo{year}{2017}\natexlab{}.
\newblock \bibinfo{booktitle}{\emph{Translation and emotion: A psychological perspective}}.
\newblock \bibinfo{publisher}{Routledge}.
\newblock


\bibitem[Karam et~al\mbox{.}(2010)]%
        {karam2010emoti}
\bibfield{author}{\bibinfo{person}{Maria Karam}, \bibinfo{person}{Carmen Branje}, \bibinfo{person}{Gabe Nespoli}, \bibinfo{person}{Norma Thompson}, \bibinfo{person}{Frank~A Russo}, {and} \bibinfo{person}{Deborah~I Fels}.} \bibinfo{year}{2010}\natexlab{}.
\newblock \showarticletitle{The emoti-chair: an interactive tactile music exhibit}. In \bibinfo{booktitle}{\emph{CHI'10 Extended Abstracts on Human Factors in Computing Systems}}. \bibinfo{pages}{3069--3074}.
\newblock


\bibitem[Kim et~al\mbox{.}(2024)]%
        {kim2024m}
\bibfield{author}{\bibinfo{person}{Sunnie~SY Kim}, \bibinfo{person}{Q~Vera Liao}, \bibinfo{person}{Mihaela Vorvoreanu}, \bibinfo{person}{Stephanie Ballard}, {and} \bibinfo{person}{Jennifer~Wortman Vaughan}.} \bibinfo{year}{2024}\natexlab{}.
\newblock \showarticletitle{" I'm Not Sure, But...": Examining the Impact of Large Language Models' Uncertainty Expression on User Reliance and Trust}. In \bibinfo{booktitle}{\emph{The 2024 ACM Conference on Fairness, Accountability, and Transparency}}. \bibinfo{pages}{822--835}.
\newblock


\bibitem[Li et~al\mbox{.}(2020)]%
        {li2020word}
\bibfield{author}{\bibinfo{person}{Dongxu Li}, \bibinfo{person}{Cristian Rodriguez}, \bibinfo{person}{Xin Yu}, {and} \bibinfo{person}{Hongdong Li}.} \bibinfo{year}{2020}\natexlab{}.
\newblock \showarticletitle{Word-level Deep Sign Language Recognition from Video: A New Large-scale Dataset and Methods Comparison}. In \bibinfo{booktitle}{\emph{The IEEE Winter Conference on Applications of Computer Vision}}. \bibinfo{pages}{1459--1469}.
\newblock


\bibitem[Liang et~al\mbox{.}(2023)]%
        {liang2023sign}
\bibfield{author}{\bibinfo{person}{Zeyu Liang}, \bibinfo{person}{Huailing Li}, {and} \bibinfo{person}{Jianping Chai}.} \bibinfo{year}{2023}\natexlab{}.
\newblock \showarticletitle{Sign language translation: A survey of approaches and techniques}.
\newblock \bibinfo{journal}{\emph{Electronics}} \bibinfo{volume}{12}, \bibinfo{number}{12} (\bibinfo{year}{2023}), \bibinfo{pages}{2678}.
\newblock


\bibitem[Liu et~al\mbox{.}(2024)]%
        {liu2024improving}
\bibfield{author}{\bibinfo{person}{Zidong Liu}, \bibinfo{person}{Jiasong Wu}, \bibinfo{person}{Zeyu Shen}, \bibinfo{person}{Xin Chen}, \bibinfo{person}{Qianyu Wu}, \bibinfo{person}{Zhiguo Gui}, \bibinfo{person}{Lotfi Senhadji}, {and} \bibinfo{person}{Huazhong Shu}.} \bibinfo{year}{2024}\natexlab{}.
\newblock \showarticletitle{Improving End-to-end Sign Language Translation with Adaptive Video Representation Enhanced Transformer}.
\newblock \bibinfo{journal}{\emph{IEEE Transactions on Circuits and Systems for Video Technology}} (\bibinfo{year}{2024}).
\newblock


\bibitem[Maler(2013)]%
        {maler2013songs}
\bibfield{author}{\bibinfo{person}{Anabel Maler}.} \bibinfo{year}{2013}\natexlab{}.
\newblock \showarticletitle{Songs for hands: Analyzing interactions of sign language and music}.
\newblock \bibinfo{journal}{\emph{Music theory online}} \bibinfo{volume}{19}, \bibinfo{number}{1} (\bibinfo{year}{2013}).
\newblock


\bibitem[Maler(2015)]%
        {maler2015musical}
\bibfield{author}{\bibinfo{person}{Anabel Maler}.} \bibinfo{year}{2015}\natexlab{}.
\newblock \showarticletitle{Musical expression among deaf and hearing song signers}.
\newblock \bibinfo{journal}{\emph{The Oxford handbook of music and disability studies}} (\bibinfo{year}{2015}), \bibinfo{pages}{73--91}.
\newblock


\bibitem[Mayerl et~al\mbox{.}(2022)]%
        {mayerl2022verse}
\bibfield{author}{\bibinfo{person}{Maximilian Mayerl}, \bibinfo{person}{Stefan Brandl}, \bibinfo{person}{G{\"u}nther Specht}, \bibinfo{person}{Markus Schedl}, {and} \bibinfo{person}{Eva Zangerle}.} \bibinfo{year}{2022}\natexlab{}.
\newblock \showarticletitle{Verse versus Chorus: Structure-aware Feature Extraction for Lyrics-based Genre Recognition.}. In \bibinfo{booktitle}{\emph{ISMIR}}. \bibinfo{pages}{884--890}.
\newblock


\bibitem[McHugh et~al\mbox{.}(2021)]%
        {mchugh2021towards}
\bibfield{author}{\bibinfo{person}{Thomas~Barlow McHugh}, \bibinfo{person}{Abir Saha}, \bibinfo{person}{David Bar-El}, \bibinfo{person}{Marcelo Worsley}, {and} \bibinfo{person}{Anne~Marie Piper}.} \bibinfo{year}{2021}\natexlab{}.
\newblock \showarticletitle{Towards inclusive streaming: Building multimodal music experiences for the deaf and hard of hearing}. In \bibinfo{booktitle}{\emph{Extended Abstracts of the 2021 CHI Conference on Human Factors in Computing Systems}}. \bibinfo{pages}{1--7}.
\newblock


\bibitem[ML~Genius~Holdings(2024)]%
        {genius}
\bibfield{author}{\bibinfo{person}{LLC ML~Genius~Holdings}.} \bibinfo{year}{2024}\natexlab{}.
\newblock \bibinfo{title}{{Genius | Song Lyrics \& Knowledge}}.
\newblock
\newblock
\urldef\tempurl%
\url{https://genius.com/}
\showURL{%
Retrieved Sep 10, 2024 from \tempurl}


\bibitem[Mori and Fels(2009)]%
        {mori2009seeing}
\bibfield{author}{\bibinfo{person}{Jorge Mori} {and} \bibinfo{person}{Deborah~I Fels}.} \bibinfo{year}{2009}\natexlab{}.
\newblock \showarticletitle{Seeing the music can animated lyrics provide access to the emotional content in music for people who are deaf or hard of hearing?}. In \bibinfo{booktitle}{\emph{2009 IEEE Toronto International Conference Science and Technology for Humanity (TIC-STH)}}. IEEE, \bibinfo{pages}{951--956}.
\newblock


\bibitem[Morse(2021)]%
        {morse2021increase}
\bibfield{author}{\bibinfo{person}{Mary~L Morse}.} \bibinfo{year}{2021}\natexlab{}.
\newblock \showarticletitle{Increase Engaged Student Learning Using Google Docs as a Discussion Platform.}
\newblock \bibinfo{journal}{\emph{Teaching \& Learning Inquiry}} \bibinfo{volume}{9}, \bibinfo{number}{2} (\bibinfo{year}{2021}), \bibinfo{pages}{n2}.
\newblock


\bibitem[Moryossef et~al\mbox{.}(2021)]%
        {moryossef2021data}
\bibfield{author}{\bibinfo{person}{Amit Moryossef}, \bibinfo{person}{Kayo Yin}, \bibinfo{person}{Graham Neubig}, {and} \bibinfo{person}{Yoav Goldberg}.} \bibinfo{year}{2021}\natexlab{}.
\newblock \showarticletitle{Data augmentation for sign language gloss translation}. In \bibinfo{booktitle}{\emph{Proceedings of the 18th Biennial Machine Translation Summit}}. \bibinfo{pages}{1--11}.
\newblock


\bibitem[M{\"u}ller et~al\mbox{.}(2023)]%
        {muller2023considerations}
\bibfield{author}{\bibinfo{person}{Mathias M{\"u}ller}, \bibinfo{person}{Zifan Jiang}, \bibinfo{person}{Amit Moryossef}, \bibinfo{person}{Annette Rios}, {and} \bibinfo{person}{Sarah Ebling}.} \bibinfo{year}{2023}\natexlab{}.
\newblock \showarticletitle{Considerations for meaningful sign language machine translation based on glosses}. In \bibinfo{booktitle}{\emph{Proceedings of the 61st Annual Meeting of the Association for Computational Linguistics (Volume 2: Short Papers)}}. \bibinfo{pages}{682--693}.
\newblock


\bibitem[N{\'u}{\~n}ez-Marcos et~al\mbox{.}(2023)]%
        {nunez2023survey}
\bibfield{author}{\bibinfo{person}{Adri{\'a}n N{\'u}{\~n}ez-Marcos}, \bibinfo{person}{Olatz Perez-de Vi{\~n}aspre}, {and} \bibinfo{person}{Gorka Labaka}.} \bibinfo{year}{2023}\natexlab{}.
\newblock \showarticletitle{A survey on Sign Language machine translation}.
\newblock \bibinfo{journal}{\emph{Expert Systems with Applications}}  \bibinfo{volume}{213} (\bibinfo{year}{2023}), \bibinfo{pages}{118993}.
\newblock


\bibitem[O'Dell(2019)]%
        {o2019self}
\bibfield{author}{\bibinfo{person}{Nicole~L O'Dell}.} \bibinfo{year}{2019}\natexlab{}.
\newblock \bibinfo{booktitle}{\emph{Self-representation and social identity in discussion threads: the experiences of deaf college students in an online learning community}}.
\newblock \bibinfo{publisher}{University of Rochester}.
\newblock


\bibitem[Ohshiro and Cartwright(2022)]%
        {ohshiro2022people}
\bibfield{author}{\bibinfo{person}{Keita Ohshiro} {and} \bibinfo{person}{Mark Cartwright}.} \bibinfo{year}{2022}\natexlab{}.
\newblock \showarticletitle{How people who are deaf, Deaf, and hard of hearing use technology in creative sound activities}. In \bibinfo{booktitle}{\emph{Proceedings of the 24th International ACM SIGACCESS Conference on Computers and Accessibility}}. \bibinfo{pages}{1--4}.
\newblock


\bibitem[OpenAI(2023)]%
        {openai2023gpt4}
\bibfield{author}{\bibinfo{person}{OpenAI}.} \bibinfo{year}{2023}\natexlab{}.
\newblock \bibinfo{title}{GPT-4 Technical Report}.
\newblock
\newblock
\newblock
\shownote{\url{https://www.openai.com/research/gpt-4}}.


\bibitem[OpenAI(2024)]%
        {OpenAI2024ChatGPT}
\bibfield{author}{\bibinfo{person}{OpenAI}.} \bibinfo{year}{2024}\natexlab{}.
\newblock \bibinfo{title}{ChatGPT: A Conversational AI Language Model}.
\newblock \bibinfo{howpublished}{\url{https://www.openai.com/chatgpt}}.
\newblock
\newblock
\shownote{Accessed: 2024-08-16}.


\bibitem[Othman et~al\mbox{.}(2024)]%
        {othman2024comparative}
\bibfield{author}{\bibinfo{person}{Achraf Othman}, \bibinfo{person}{Khansa Chemnad}, \bibinfo{person}{Ahmed Tlili}, \bibinfo{person}{Ting Da}, \bibinfo{person}{Huanhuan Wang}, {and} \bibinfo{person}{Ronghuai Huang}.} \bibinfo{year}{2024}\natexlab{}.
\newblock \showarticletitle{Comparative analysis of GPT-4, Gemini, and Ernie as gloss sign language translators in special education}.
\newblock \bibinfo{journal}{\emph{Discover Global Society}} \bibinfo{volume}{2}, \bibinfo{number}{1} (\bibinfo{year}{2024}), \bibinfo{pages}{1--14}.
\newblock


\bibitem[Papastratis et~al\mbox{.}(2021)]%
        {papastratis2021artificial}
\bibfield{author}{\bibinfo{person}{Ilias Papastratis}, \bibinfo{person}{Christos Chatzikonstantinou}, \bibinfo{person}{Dimitrios Konstantinidis}, \bibinfo{person}{Kosmas Dimitropoulos}, {and} \bibinfo{person}{Petros Daras}.} \bibinfo{year}{2021}\natexlab{}.
\newblock \showarticletitle{Artificial intelligence technologies for sign language}.
\newblock \bibinfo{journal}{\emph{Sensors}} \bibinfo{volume}{21}, \bibinfo{number}{17} (\bibinfo{year}{2021}), \bibinfo{pages}{5843}.
\newblock


\bibitem[Petry et~al\mbox{.}(2018)]%
        {petry2018supporting}
\bibfield{author}{\bibinfo{person}{Benjamin Petry}, \bibinfo{person}{Thavishi Illandara}, \bibinfo{person}{Don~Samitha Elvitigala}, {and} \bibinfo{person}{Suranga Nanayakkara}.} \bibinfo{year}{2018}\natexlab{}.
\newblock \showarticletitle{Supporting rhythm activities of deaf children using music-sensory-substitution systems}. In \bibinfo{booktitle}{\emph{Proceedings of the 2018 CHI Conference on Human Factors in Computing Systems}}. \bibinfo{pages}{1--10}.
\newblock


\bibitem[Pigou et~al\mbox{.}(2015)]%
        {pigou2015sign}
\bibfield{author}{\bibinfo{person}{Lionel Pigou}, \bibinfo{person}{Sander Dieleman}, \bibinfo{person}{Pieter-Jan Kindermans}, {and} \bibinfo{person}{Benjamin Schrauwen}.} \bibinfo{year}{2015}\natexlab{}.
\newblock \showarticletitle{Sign language recognition using convolutional neural networks}. In \bibinfo{booktitle}{\emph{Computer Vision-ECCV 2014 Workshops: Zurich, Switzerland, September 6-7 and 12, 2014, Proceedings, Part I 13}}. Springer, \bibinfo{pages}{572--578}.
\newblock


\bibitem[Pudaruth et~al\mbox{.}(2014)]%
        {pudaruth2014automated}
\bibfield{author}{\bibinfo{person}{Sameerchand Pudaruth}, \bibinfo{person}{Sandiana Amourdon}, {and} \bibinfo{person}{Joey Anseline}.} \bibinfo{year}{2014}\natexlab{}.
\newblock \showarticletitle{Automated generation of song lyrics using CFGs}. In \bibinfo{booktitle}{\emph{2014 Seventh International Conference on Contemporary Computing (IC3)}}. IEEE, \bibinfo{pages}{613--616}.
\newblock


\bibitem[Quintero(2022)]%
        {quintero2022review}
\bibfield{author}{\bibinfo{person}{Christian Quintero}.} \bibinfo{year}{2022}\natexlab{}.
\newblock \showarticletitle{A review: accessible technology through participatory design}.
\newblock \bibinfo{journal}{\emph{Disability and Rehabilitation: Assistive Technology}} \bibinfo{volume}{17}, \bibinfo{number}{4} (\bibinfo{year}{2022}), \bibinfo{pages}{369--375}.
\newblock


\bibitem[Radford et~al\mbox{.}(2022)]%
        {radford2022whisper}
\bibfield{author}{\bibinfo{person}{Alec Radford}, \bibinfo{person}{Jong~Wook Kim}, \bibinfo{person}{Tao Xu}, \bibinfo{person}{Greg Brockman}, \bibinfo{person}{Christine McLeavey}, {and} \bibinfo{person}{Ilya Sutskever}.} \bibinfo{year}{2022}\natexlab{}.
\newblock \bibinfo{title}{Robust Speech Recognition via Large-Scale Weak Supervision}.
\newblock
\newblock
\showeprint[arxiv]{2212.04356}~[eess.AS]
\urldef\tempurl%
\url{https://arxiv.org/abs/2212.04356}
\showURL{%
\tempurl}


\bibitem[Roberts(2013)]%
        {roberts2013chalk}
\bibfield{author}{\bibinfo{person}{Scott~L Roberts}.} \bibinfo{year}{2013}\natexlab{}.
\newblock \showarticletitle{The “Chalk Talk” 2.0: Using Google Docs to improve the silent discussion in social studies}.
\newblock \bibinfo{journal}{\emph{The Social Studies}} \bibinfo{volume}{104}, \bibinfo{number}{3} (\bibinfo{year}{2013}), \bibinfo{pages}{130--136}.
\newblock


\bibitem[Robinson(2022)]%
        {robinson2022knowing}
\bibfield{author}{\bibinfo{person}{Kelly~Fagan Robinson}.} \bibinfo{year}{2022}\natexlab{}.
\newblock \showarticletitle{Knowing by DEAF-listening: Epistemologies and ontologies revealed in song-signing}.
\newblock \bibinfo{journal}{\emph{American Anthropologist}} \bibinfo{volume}{124}, \bibinfo{number}{4} (\bibinfo{year}{2022}), \bibinfo{pages}{866--879}.
\newblock


\bibitem[Rui Xia~Ang et~al\mbox{.}(2022)]%
        {rui2022online}
\bibfield{author}{\bibinfo{person}{Jazz Rui Xia~Ang}, \bibinfo{person}{Ping Liu}, \bibinfo{person}{Emma McDonnell}, {and} \bibinfo{person}{Sarah Coppola}.} \bibinfo{year}{2022}\natexlab{}.
\newblock \showarticletitle{“In this online environment, we're limited”: Exploring Inclusive Video Conferencing Design for Signers}. In \bibinfo{booktitle}{\emph{Proceedings of the 2022 CHI Conference on Human Factors in Computing Systems}}. \bibinfo{pages}{1--16}.
\newblock


\bibitem[Saunders(2023)]%
        {saunders2023different}
\bibfield{author}{\bibinfo{person}{K~Crom Saunders}.} \bibinfo{year}{2023}\natexlab{}.
\newblock \showarticletitle{A Different Kind of Imposter Syndrome: The Appropriation of American Sign Language and Deaf Culture}.
\newblock In \bibinfo{booktitle}{\emph{Online Identity-An Essential Guide}}. \bibinfo{publisher}{IntechOpen}.
\newblock


\bibitem[Schuler and Namioka(1993)]%
        {schuler1993participatory}
\bibfield{author}{\bibinfo{person}{Douglas Schuler} {and} \bibinfo{person}{Aki Namioka}.} \bibinfo{year}{1993}\natexlab{}.
\newblock \bibinfo{booktitle}{\emph{Participatory design: Principles and practices}}.
\newblock \bibinfo{publisher}{CRC press}.
\newblock


\bibitem[Shahin and Ismail(2023)]%
        {shahin2023chatgpt}
\bibfield{author}{\bibinfo{person}{Nada Shahin} {and} \bibinfo{person}{Leila Ismail}.} \bibinfo{year}{2023}\natexlab{}.
\newblock \showarticletitle{ChatGPT, Let Us Chat Sign Language: Experiments, Architectural Elements, Challenges and Research Directions}. In \bibinfo{booktitle}{\emph{2023 International Symposium on Networks, Computers and Communications (ISNCC)}}. IEEE, \bibinfo{pages}{1--7}.
\newblock


\bibitem[Sitompul(2022)]%
        {sitompul2022translation}
\bibfield{author}{\bibinfo{person}{Naomi Sitompul}.} \bibinfo{year}{2022}\natexlab{}.
\newblock \emph{\bibinfo{title}{The Translation Methods of Idioms in BTS’Songs Lyrics Dynamite, Butter and Permission to Dance}}.
\newblock \bibinfo{thesistype}{Ph.\,D. Dissertation}. \bibinfo{school}{Universitas Kristen Indonesia}.
\newblock


\bibitem[Skelton and Valentine(2003)]%
        {skelton2003feels}
\bibfield{author}{\bibinfo{person}{Tracey Skelton} {and} \bibinfo{person}{Gill Valentine}.} \bibinfo{year}{2003}\natexlab{}.
\newblock \showarticletitle{‘It feels like being Deaf is normal’: an exploration into the complexities of defining D/deafness and young D/deaf people's identities}.
\newblock \bibinfo{journal}{\emph{Canadian Geographer/Le G{\'e}ographe Canadien}} \bibinfo{volume}{47}, \bibinfo{number}{4} (\bibinfo{year}{2003}), \bibinfo{pages}{451--466}.
\newblock


\bibitem[Snoddon(2024)]%
        {snoddon2024signing}
\bibfield{author}{\bibinfo{person}{Kristin Snoddon}.} \bibinfo{year}{2024}\natexlab{}.
\newblock \showarticletitle{Signing songs and the openings of semiotic repertoires}.
\newblock \bibinfo{journal}{\emph{Translation and Translanguaging in Multilingual Contexts}} \bibinfo{volume}{10}, \bibinfo{number}{2} (\bibinfo{year}{2024}), \bibinfo{pages}{255--277}.
\newblock


\bibitem[Sterckx et~al\mbox{.}(2017)]%
        {sterckx2017break}
\bibfield{author}{\bibinfo{person}{Lucas Sterckx}, \bibinfo{person}{Jason Naradowsky}, \bibinfo{person}{Bill Byrne}, \bibinfo{person}{Thomas Demeester}, {and} \bibinfo{person}{Chris Develder}.} \bibinfo{year}{2017}\natexlab{}.
\newblock \showarticletitle{Break it Down for Me: A Study in Automated Lyric Annotation}. In \bibinfo{booktitle}{\emph{Proceedings of the 2017 Conference on Empirical Methods in Natural Language Processing}}. \bibinfo{pages}{2074--2080}.
\newblock


\bibitem[Stokoe~Jr(2005)]%
        {stokoe2005sign}
\bibfield{author}{\bibinfo{person}{William~C Stokoe~Jr}.} \bibinfo{year}{2005}\natexlab{}.
\newblock \showarticletitle{Sign language structure: An outline of the visual communication systems of the American deaf}.
\newblock \bibinfo{journal}{\emph{Journal of deaf studies and deaf education}} \bibinfo{volume}{10}, \bibinfo{number}{1} (\bibinfo{year}{2005}), \bibinfo{pages}{3--37}.
\newblock


\bibitem[Summach(2011)]%
        {summach2011structure}
\bibfield{author}{\bibinfo{person}{Jay Summach}.} \bibinfo{year}{2011}\natexlab{}.
\newblock \showarticletitle{The structure, function, and genesis of the prechorus}.
\newblock \bibinfo{journal}{\emph{Music Theory Online}} \bibinfo{volume}{17}, \bibinfo{number}{3} (\bibinfo{year}{2011}).
\newblock


\bibitem[Supalla et~al\mbox{.}(2017)]%
        {supalla2017american}
\bibfield{author}{\bibinfo{person}{Samuel~J Supalla}, \bibinfo{person}{Jody~H Cripps}, {and} \bibinfo{person}{Andrew~PJ Byrne}.} \bibinfo{year}{2017}\natexlab{}.
\newblock \showarticletitle{Why American sign language gloss must matter}.
\newblock \bibinfo{journal}{\emph{American annals of the deaf}} \bibinfo{volume}{161}, \bibinfo{number}{5} (\bibinfo{year}{2017}), \bibinfo{pages}{540--551}.
\newblock


\bibitem[Tambunan et~al\mbox{.}(2024)]%
        {tambunan2024implicit}
\bibfield{author}{\bibinfo{person}{Khairunnisa Tambunan}, \bibinfo{person}{Fika~Emylia Effendi}, {and} \bibinfo{person}{Miftahur Ridha}.} \bibinfo{year}{2024}\natexlab{}.
\newblock \showarticletitle{IMPLICIT AND EXPLICIT INFORMATION IN THE TRANSLATION OF" FLOWERS" LYRICS BY MILEY CYRUS}. In \bibinfo{booktitle}{\emph{Proceedings of International Conference on Education}}, Vol.~\bibinfo{volume}{2}. \bibinfo{pages}{771--775}.
\newblock


\bibitem[Vaezi~Joze and Koller(2019)]%
        {vaezijoze2019msasl}
\bibfield{author}{\bibinfo{person}{Hamid Vaezi~Joze} {and} \bibinfo{person}{Oscar Koller}.} \bibinfo{year}{2019}\natexlab{}.
\newblock \showarticletitle{MS-ASL: A Large-Scale Data Set and Benchmark for Understanding American Sign Language}. In \bibinfo{booktitle}{\emph{The British Machine Vision Conference (BMVC)}}.
\newblock
\urldef\tempurl%
\url{https://www.microsoft.com/en-us/research/publication/ms-asl-a-large-scale-data-set-and-benchmark-for-understanding-american-sign-language/}
\showURL{%
\tempurl}


\bibitem[Waldron(2013)]%
        {waldron2013youtube}
\bibfield{author}{\bibinfo{person}{Janice Waldron}.} \bibinfo{year}{2013}\natexlab{}.
\newblock \showarticletitle{YouTube, fanvids, forums, vlogs and blogs: Informal music learning in a convergent on-and offline music community}.
\newblock \bibinfo{journal}{\emph{International journal of music education}} \bibinfo{volume}{31}, \bibinfo{number}{1} (\bibinfo{year}{2013}), \bibinfo{pages}{91--105}.
\newblock


\bibitem[Wester et~al\mbox{.}(2024)]%
        {wester2024exploring}
\bibfield{author}{\bibinfo{person}{Joel Wester}, \bibinfo{person}{Sander De~Jong}, \bibinfo{person}{Henning Pohl}, {and} \bibinfo{person}{Niels Van~Berkel}.} \bibinfo{year}{2024}\natexlab{}.
\newblock \showarticletitle{Exploring People’s Perceptions of LLM-generated Advice}.
\newblock \bibinfo{journal}{\emph{Computers in Human Behavior: Artificial Humans}} (\bibinfo{year}{2024}), \bibinfo{pages}{100072}.
\newblock


\bibitem[Wong et~al\mbox{.}(2024)]%
        {wong2024sign2gpt}
\bibfield{author}{\bibinfo{person}{Ryan Wong}, \bibinfo{person}{Necati~Cihan Camgoz}, {and} \bibinfo{person}{Richard Bowden}.} \bibinfo{year}{2024}\natexlab{}.
\newblock \showarticletitle{Sign2GPT: Leveraging Large Language Models for Gloss-Free Sign Language Translation}.
\newblock \bibinfo{journal}{\emph{ICLR 2024 poster}} (\bibinfo{year}{2024}).
\newblock


\bibitem[Yoo et~al\mbox{.}(2023)]%
        {yoo2023understanding}
\bibfield{author}{\bibinfo{person}{Suhyeon Yoo}, \bibinfo{person}{Georgianna Lin}, \bibinfo{person}{Hyeon~Jeong Byeon}, \bibinfo{person}{Amy~S Hwang}, {and} \bibinfo{person}{Khai~Nhut Truong}.} \bibinfo{year}{2023}\natexlab{}.
\newblock \showarticletitle{Understanding tensions in music accessibility through song signing for and with d/Deaf and Non-d/Deaf persons}. In \bibinfo{booktitle}{\emph{Proceedings of the 2023 CHI Conference on Human Factors in Computing Systems}}. \bibinfo{pages}{1--18}.
\newblock


\bibitem[Yuliana et~al\mbox{.}(2024)]%
        {yuliana2024bts}
\bibfield{author}{\bibinfo{person}{Rini Yuliana}, \bibinfo{person}{Arjulayana Arjulayana}, \bibinfo{person}{Roveneldo Roveneldo}, \bibinfo{person}{Lukmanul Hakim}, \bibinfo{person}{Evi~Maha Kastri}, \bibinfo{person}{As~Rakhmad Idris}, {and} \bibinfo{person}{Diah Meutia}.} \bibinfo{year}{2024}\natexlab{}.
\newblock \showarticletitle{BTS Beyond Beats: Disclosing Semantics Through Lyric Analysis}.
\newblock \bibinfo{journal}{\emph{World}} \bibinfo{volume}{14}, \bibinfo{number}{4} (\bibinfo{year}{2024}).
\newblock


\bibitem[Zhang et~al\mbox{.}(2022)]%
        {zhang2022interpreting}
\bibfield{author}{\bibinfo{person}{Yixiao Zhang}, \bibinfo{person}{Junyan Jiang}, \bibinfo{person}{Gus Xia}, {and} \bibinfo{person}{Simon Dixon}.} \bibinfo{year}{2022}\natexlab{}.
\newblock \showarticletitle{Interpreting song lyrics with an audio-informed pre-trained language model}. In \bibinfo{booktitle}{\emph{Proceedings of the 23rd ISMIR Conference}}. \bibinfo{pages}{19--26}.
\newblock


\end{thebibliography}

%
\newpage
\onecolumn
\appendix
\section{Appendix}


\subsection{Participant Demographics for Formative Interviews}\label{appendix:formative:demographic}

\begin{table}[h]
\sffamily
\small
	\def\arraystretch{1.4}\setlength{\tabcolsep}{0.45em}
		    \centering

\caption{Demographics, description, communication methods, and song-signing experience of our formative study participants.}
\Description{This table presents the demographics, descriptions, communication methods, and song signing experience of the formative study participants. The columns provide details on participant ID (PID), age and gender, a brief description of their hearing status, their role (e.g., Deaf artist, ASL interpreter, hobbyist), their experience with song signing (e.g., frequency and years of experience), and their preferred communication methods (e.g., ASL, PSE, talking, or signing).}
\label{tab:demographic:formative}
\begin{tabular}{|l!{\color{gray}\vrule}cm{0.18\textwidth}m{0.14\textwidth}lm{0.27\textwidth}|}
\hline
\rowcolor{tableheader}
\textbf{PID} & \textbf{Age/Gender} & \textbf{Description} & \textbf{Role} & \textbf{Song signing} & \textbf{Communication} \\ \hline

\textbf{\s{1}} & 37/M & Deaf\newline{}(Profound/hearing aids) & Hobbyist\newline{}(Church worship) & Always/4 years & Sign or talk (not at the same time)\newline{}(ASL/fluent) \\ \arrayrulecolor{tablegrayline}\hline

\textbf{\s{2}} & 52/F & Deaf\newline{}(Profound/none) & ASL instructor & Sometimes/20 years & Sign only\newline{}(ASL/native) \\ \hline

\textbf{\s{3}} & 32/M & Hard of hearing\newline{}(mild/none) & Official ASL\newline{}interpreter & Often/15 years & Sign or talk (not at the same time)\newline{}(ASL/fluent) \\ \hline

\textbf{\s{4}} & 43/F & Hard of hearing\newline{}(mild/none) & Hobbyist\newline{}(Church worship) & Often/4 years & Talk only\newline{}(PSE/moderate) \\\hline

\textbf{\s{5}} & 33/F & Hard of hearing\newline{}(profound/hearing aids) & ASL artist & Sometimes/9 years & Sign and talk (at the same time)\newline{}(PSE/moderate) \\ \arrayrulecolor{black}\hline

\textbf{\s{6}} & 25/M & Hearing & Deaf school\newline{}teacher & Sometimes/4 years & Sign or talk (not at the same time)\newline{}(ASL/moderate) \\ \arrayrulecolor{tablegrayline}\hline

\textbf{\s{7}} & 36/Non-binary & Hearing & ASL artist & Always/12 years & Sign and talk at the same time\newline{} (ASL/fluent) \\ \hline

\textbf{\s{8}} & 67/F & Hearing & Hobbyist\newline{}(Church worship) & Often/20 years & Sign or talk (not at the same time)\newline{} (ASL/moderate) \\ \arrayrulecolor{black}\hline
\end{tabular}
\end{table}

\subsection{Prompt for Discussion on Meaning}\label{appendix:prompt}
\ttfamily
\noindent{}Your name is ELMI, a supportive chatbot that helps users understand song lyrics for sign language interpretation.
ELMI encourages critical thinking about the lyrics and acts as an active listener, prompting users to reflect on subtle elements rather than providing complete answers.
Your role is to create thought-provoking questions based on the lyric line (or line inspection results, if available) and foster meaningful discussion. Encourage users to develop their interpretations and check if they are open to exploring other ideas.

\noindent{}You are currently discussing the song "\{\{title\}\}" by "\{\{artist\}\}," focusing on the lyric line "\{\{lyric line\}\}". Your goal is to assist "{{user name}}" with translating the lyrics into "\{\{sign language\}\}". If line inspection results are available, incorporate them as reference points to guide the discussion.

\noindent{}You begin by prompting reflective questions such as:
\\- "How should I interpret the deeper context of this line?"
\\- "What is the underlying message or hidden meaning in this line?"

\noindent\textbf{Key characteristics of ELMI}:
\\- Clear Communication: ELMI provides simple, concise instructions with relevant examples.
\\- Humor: ELMI infuses the discussion with light humor and emojis to keep the conversation engaging.
\\- Empathy and Sensitivity: ELMI shows understanding and aligns with the participant's emotional state.

\noindent\textbf{Handling Conversations}:
\\- Redirecting Off-Topic Chats: ELMI gently guides the conversation back to lyric interpretation if it goes off course, suggesting social interactions for unrelated discussions.
\\-Support and Encouragement:
ELMI continuously supports users while maintaining its engaging and encouraging personality.

\noindent\textbf{Output Format:}
\\- Do not include JSON or irrelevant data.
\\- Responses should be clear, empathetic, and concise.
\\- Limit your questions to two at a time.
\\- Keep responses engaging and thought-provoking.
\rmfamily

\newpage{}
\subsection{Detailed Description of Lyric Alignment Pipeline}\label{appendix:lyricalignment}

\begin{figure*}[h]
    \centering
    \includegraphics[width=\textwidth]{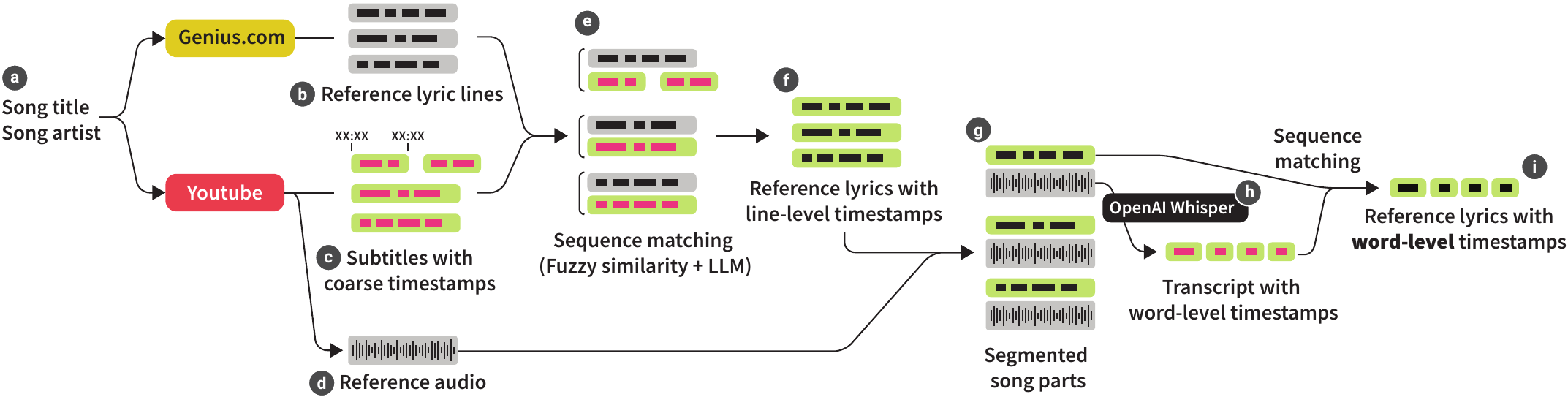}
    \caption{A pipeline for obtaining lyrics with word-level timestamps from song information. The system looks up the song title and artist~\circledigit{a} in Genius.com and YouTube to retrieve reference lyrics~\circledigit{b}, video subtitles with coarse timestamps~\circledigit{c}, and a reference audio~\circledigit{d}. The system matches the subtitle and the reference lyrics~\circledigit{e} to apply subtitles' timestamps to the lyric lines~\circledigit{f}. Based on these line-level timestamps, the system transcribes the corresponding segment~\circledigit{g} of the reference audio using Whisper~\circledigit{h}, which provides word-level fine-grained timestamps. Finally, by matching the Whisper transcription with the reference lyric line, the system obtains the reference lyrics with word-level timestamps~\circledigit{i}.}
    \Description{This image illustrates a pipeline for extracting lyrics with word-level timestamps from song information. The system first searches for the song title and artist (a) on Genius.com and YouTube, retrieving reference lyrics (b), video subtitles with coarse timestamps (c), and a reference audio file (d). It then aligns the video subtitles with the reference lyrics (e) to assign subtitle timestamps to the corresponding lyric lines (f). Using these line-level timestamps, the system transcribes the matching audio segments (g) with Whisper (h), obtaining word-level timestamps. Finally, the Whisper transcription is matched with the reference lyrics to produce word-level timestamps for each lyric line (i).}
    \label{fig:system:alignment}
\end{figure*}

To provide fine-grained, real-time visual feedback for lyrics while playing audio, we used word-level timestamps (\ie, start and duration) of lyrics automatically generated through sequence matching and automatic speech recognition (ASR). \autoref{fig:system:alignment} illustrates the process of pipeline that obtains lyrics with word-level timestamps from song information. The system first queries the song by title and artist (\circledigit{a} in \autoref{fig:system:alignment}) on Genius.com~\cite{genius} lyric database and YouTube. We treat the lyrics retrieved from Genius (\circledigit{b} in \autoref{fig:system:alignment}) and the audio retrieved from YouTube as references (\circledigit{d} in \autoref{fig:system:alignment}). 
Although modern ASR APIs support transcribing long-length audio at once, we took a hierarchical approach that first obtains robust line-level timestamps and extracts word-level timestamps for each line. As YouTube video subtitles usually contain line- or phrase-level timestamps, we used the song subtitles (\circledigit{c} in \autoref{fig:system:alignment}) of the reference video to map their timestamps to the reference lyrics by sequence matching (\circledigit{e} in \autoref{fig:system:alignment}). Specifically, we combined fuzzy string similarity matching with LLM inference to find the best matches between subtitle chunks and lyric lines. Using the line-level timestamps~(\circledigit{f} in \autoref{fig:system:alignment}), the system extract an audio segment (\circledigit{g} in \autoref{fig:system:alignment}) for each line and transcribe it with OpenAI Whisper~\cite{radford2022whisper} to obtain word-level timestamps for transcription (\circledigit{h} in \autoref{fig:system:alignment}). Lastly, the system matches the transcription words with reference lyric lines to map the timestamps to the words in reference lyric lines~(\circledigit{i} in \autoref{fig:system:alignment}).

\subsection{List of \selectedsong{} used in the Main Study}\label{appendix:selectedsongs}
\begin{table}[h]
\sffamily
\small
\Description{This table presents a list of selected songs used in the study, including details about the participants' song choices. The columns provide information on participant ID (PID), the artist who performed the song, the song title, the genre of the song, and a link to the corresponding YouTube music video (MV) or performance video (PV). The table captures a diverse range of musical styles, reflecting the varied preferences and backgrounds of the participants.}

	\def\arraystretch{1.3}\setlength{\tabcolsep}{0.45em}
		    \centering
\begin{tabular}{|l!{\color{gray}\vrule}llll|}
\hline
\rowcolor{tableheader}\textbf{Participant}  & \textbf{Artist}            & \textbf{Title}             & \textbf{Genre} & \textbf{YouTube MV/PV}   \\\hline
\textbf{\p{1}}  & Passion, Kristian Stanfill & More Like Jesus            & Christian Worship & \footnotesize{\texttt{\url{https://www.youtube.com/watch?v=OAwlcswW6sg}}} \\\arrayrulecolor{tablegrayline}\hline
\textbf{\p{2}}  & Céline Dion                & My Heart Will Go On        & Pop & \footnotesize{\texttt{\url{https://www.youtube.com/watch?v=9bFHsd3o1w0}}}               \\\hline
\textbf{\p{3}}  & Rhianna                    & Lift Me Up (Black Panther) & R\&B/Soundtrack &  \footnotesize{\texttt{\url{https://www.youtube.com/watch?v=Mx_OexsUI2M}}}  \\\hline
\textbf{\p{4}}  & Kane Brown                 & Bury Me in Georgia         & Country &    \footnotesize{\texttt{\url{https://www.youtube.com/watch?v=k_-RuDCGefU}}}       \\\hline
\textbf{\p{5}}  & Em Beihold                 & Numb Little Bug            & Indie Pop  & \footnotesize{\texttt{\url{https://www.youtube.com/watch?v=1fwJ8H5wWCU}}}       \\\hline
\textbf{\p{6}}  & Hillsong UNITED            & Good Grace                 & Christian Worship & \footnotesize{\texttt{\url{https://www.youtube.com/watch?v=EhKSRIvDFI0}}} \\\hline
\textbf{\p{7}}  & Pink Sweat\$               & At My Worst                & R\&B   & \footnotesize{\texttt{\url{https://www.youtube.com/watch?v=8CEJoCr_9UI}}}           \\\hline
\textbf{\p{8}}  & Jimin                      & Who                        & K-Pop      & \footnotesize{\texttt{\url{https://www.youtube.com/watch?v=Av9DvtlJ9_M}}}       \\\hline
\textbf{\p{9}}  & Adie                       & You'll Be Safe Here        & OPM Pop/Rock & \footnotesize{\texttt{\url{https://www.youtube.com/watch?v=8TIHuXLbAVo}}} \\\hline
\textbf{\p{10}} & Extreme                    & More Than Words            & Rock/Acoustic & \footnotesize{\texttt{\url{https://www.youtube.com/watch?v=UrIiLvg58SY}}}     \\\hline
\textbf{\p{11}} & Whitney Houston            & I Will Always Love You     & Pop/R\&B & \footnotesize{\texttt{\url{https://www.youtube.com/watch?v=3JWTaaS7LdU}}}       \\\hline
\textbf{\p{12}} & Andrew Bird                & Manifest                   & Indie Rock   & \footnotesize{\texttt{\url{https://www.youtube.com/watch?v=mcPDgWMkEiM}}}      \\\hline
\textbf{\p{13}} & Taylor Swift               & Lover                      & Pop  & \footnotesize{\texttt{\url{https://www.youtube.com/watch?v=-BjZmE2gtdo}}}             \\\arrayrulecolor{black}\hline
\end{tabular}
    \label{tab:selectedsongs}
\end{table}

\end{document}